\documentclass[aps,physrev,reprint,twocolumn,nofootinbib,superscriptaddress,floatfix,amsmath,amssymb]{revtex4-1}

\usepackage{graphicx} 
\usepackage{float}
\usepackage[T1]{fontenc}
\usepackage[english]{babel}
\usepackage{microtype}

\usepackage{textgreek}

\usepackage[version=4]{mhchem}
\usepackage{chemmacros}

\usepackage[separate-uncertainty=true,per-mode=symbol, detect-all]{siunitx} 
\newcommand{\nm}{\,\si{\nano\meter}}
\newcommand{\um}{\,\si{\micro\meter}}

\newcommand{\mM}{\,\si{\milli\textsc{m}}}

\usepackage{tabularx}
\usepackage{booktabs}
\AtBeginDocument{%
\heavyrulewidth=.08em
\lightrulewidth=.05em
\cmidrulewidth=.03em
\belowrulesep=.65ex
\belowbottomsep=0pt
\aboverulesep=.4ex
\abovetopsep=0pt
\cmidrulesep=\doublerulesep
\cmidrulekern=.5em
\defaultaddspace=.5em
}

\usepackage[hidelinks]{hyperref}
\usepackage{cleveref}

\crefname{equation}{Eq.}{Eqs.}
\crefname{figure}{Fig.}{Figs.}
\crefrangeformat{equation}{Eqs.~#3(#1)#4--#5(#2)#6}

\usepackage[a4paper,margin=1.5cm]{geometry}

\newcommand{\affleiden}{\affiliation{\small Huygens-Kamerlingh Onnes Laboratory, Leiden University, Leiden, The Netherlands}}
\newcommand{\affutchem}{\affiliation{\small Debye Institute for Nanomaterials research, Department of Chemistry, Utrecht University, Utrecht, The Netherlands}}
\newcommand{\affutphys}{\affiliation{\small Debye Institute for Nanomaterials research, Department of Physics, Utrecht University, Utrecht, The Netherlands}}
\newcommand{\affindia}{\affiliation{\small Department of Physics, Birla Institute of Technology and Science, Pilani - K K Birla Goa Campus, Zuarinagar, Goa, India}}
\newcommand{\affchina}{\affiliation{\small Academy of Advanced Optoelectronics, South China Normal University, Guangzhou, China}}

\begin{document}

\title{{\Large\bf Conformations and diffusion of flexibly linked colloidal chains}}

\author{Ruben W. Verweij} 
\affleiden
\author{Pepijn G. Moerman}
\affutphys
\affutchem
\author{Loes P. P. Huijnen}
\affleiden
\author{Nathalie E. G. Ligthart}
\affutchem
\author{Indrani Chakraborty}
\affleiden
\affindia
\author{Jan Groenewold}
\affutchem
\affchina
\author{Willem K. Kegel}
\affutchem
\author{Alfons van Blaaderen}
\affutphys
\author{Daniela J. Kraft}
\email[Corresponding email: ]{kraft@physics.leidenuniv.nl}
\affleiden

\date{\today}

\defcitealias{Verweij2020}{Verweij~and~Moerman~\textit{et al.}~\cite{Verweij2020}}

\begin{abstract}

For biologically relevant macromolecules such as intrinsically disordered
proteins, internal degrees of freedom that allow for shape changes have a large
influence on both the motion and function of the compound. A detailed
understanding of the effect of flexibility is needed in order to explain their
behavior. Here, we study a model system of freely-jointed chains of three to
six colloidal spheres, using both simulations and experiments. We find that in
spite of their short lengths, their conformational statistics are well
described by two dimensional Flory theory, while their average translational
and rotational diffusivity follow the Kirkwood-Riseman scaling. Their maximum
flexibility does not depend on the length of the chain, but is determined by
the near-wall in-plane translational diffusion coefficient of an individual
sphere.  Furthermore, we uncover shape-dependent effects in the short-time
diffusivity of colloidal tetramer chains, as well as non-zero couplings between
the different diffusive modes. Our findings may have implications for
understanding both the diffusive behavior and the most likely conformations of
macromolecular systems in biology and industry, such as proteins, polymers,
single-stranded DNA and other chain-like molecules.  

\end{abstract}

\maketitle

\section{Introduction}

For biologically relevant (macro)molecules, internal degrees
of freedom that allow for shape changes have a large influence on both the
motion and function of the
compound~\cite{Serdyuk2007,Torre1994,Gregory1987,Yguerabide1970,Illien2017}.
Examples of flexible systems found in nature include bio-polymers such as DNA
and transfer RNA \cite{Mellado1988}, antibodies
\cite{Yguerabide1970,Torre1994,Burton1987,Barbato1992} and intrinsically
disordered proteins (IDPs) \cite{dyson2005intrinsically,Ishino2014}. IDPs in
particular display large shape changes, due to unstructured (and therefore
flexible) regions of small hydrophilic units that typically function as linkers
between more structured domains. These are involved in important cellular
processes such as signaling and transcription. Additionally, they are often
involved in disease-related gene truncations or translocations. The coupled
binding and folding of these flexible regions lead to a large number of
possible interactions within the same set of
proteins~\cite{dyson2005intrinsically}. Even for more rigid proteins, shape
changes can be an important factor, for example in protein breathing, that
involves slow collective movements of larger secondary
structures~\cite{makowski2008molecular}. Therefore, quantitative knowledge of
structural flexibility is necessary to understand the transport properties and
function of flexible biopolymers and proteins \cite{Rundqvist2009}.

Measuring molecular shape changes calls for single-molecule techniques with a
simultaneously high spatial and temporal resolution. As a simpler alternative,
colloidal particles have been widely used as model systems for (macro)molecular
structures since the early 20th
century~\cite{Einstein1905a,Sutherland1905,Perrin1909}, because of their unique
combination of microscopic size and sensitivity to thermal fluctuations.
Although colloidal particles are frequently used as model systems, the study of
colloidal particles is interesting in its own right, as colloids can in
principle form the building blocks of materials with novel properties, such as
photonic bandgap materials~\cite{he2020colloidal}. Colloidal structures of
reconfigurable shape are expected to aid in the assembly of these structures,
because they allow the formed materials to quickly relax towards their
thermodynamic equilibrium configuration, thereby mitigating the random parking
problem~\cite{mansfield1996random}. In addition, they could provide ways to
build switchable materials~\cite{mihut2017assembling}. 

While the Brownian motion of rigid colloids of various shapes has been
extensively studied, for example for
ellipsoids~\cite{Han2006a,Meunier1994,Zahn1994},
boomerangs~\cite{Chakrabarty2014,Chakrabarty2016,Koens2017}, and
clusters~\cite{Kraft2013a,Fung2013}, most compounds found in nature show some
degree of flexibility, which may affect their transport properties. It was
proposed to calculate their diffusive properties in an approximate way, by
treating the structure as an (instantaneously) rigid body and take the ensemble
average of all possible `snapshots' of conformations, the so-called rigid body
approximation~\cite{riseman1950intrinsic,Torre1994,Iniesta1988,GarciadelaTorre2002}.
However, the accuracy of this approximation is as of yet unclear: importantly,
deviations between this approximation and the real transport properties can
become larger as function of the flexibility of the
molecule~\cite{schmidt2012translational}. 

Recently, we have studied both numerically and experimentally the effect of
segmental flexibility~\cite{Verweij2020} in a simple model system consisting of
a freely-jointed chain of three spherical colloidal particles, also called
trimers or trumbbells~\cite{Harvey1983,Roitman2005}. This was made possible for
the first time thanks to the development of colloidal structures with
freely-jointed bonds~\cite{VanDerMeulen2013}. Similar to rigid particles, we
found that shape affects the diffusive motion of the colloid at short
timescales and that displacements are larger in directions that correspond to
smaller hydrodynamic drag. By comparing our flexible trimers to rigid ones, we
found that the flexibility of the trimers led to a higher translational
diffusion coefficient. Furthermore, we uncovered a Brownian quasiscallop mode,
where diffusive motion is coupled to Brownian shape changes.  At longer
timescales, in addition to the rotational diffusion time, an analogous
conformational diffusion time governs the relaxation of the diffusive motion,
unique to flexible assemblies~\cite{Verweij2020}. These effects taken together
show that the rigid body approximation is not sufficient to model the rich
behavior of flexible objects. However, in the case of long polymer chains,
Kirkwood-Riseman theory~\cite{riseman1950intrinsic}, which is based on the
rigid body approximation, is able to at least describe equilibrium properties
such as the average translational diffusion
coefficient~\cite{yamada1992transport}. This is attractive because, if
accurate, it would provide a simple and quick way to calculate the equilibrium
long time diffusion coefficients typically measured in light scattering
experiments.

Here, we study flexible chains of three to six particles using both experiments
and simulations. Conceptually, the longer chains most resemble a flexible
polymer, modeled by beads on a chain, while the shorter chains are expected to
show deviations from predictions based on polymer theory. We set out to test to
what extent the conformations of our bead chains can be described by polymer
theory and to what extent their equilibrium diffusivity can be described by
simple scalings such as the Kirkwood-Riseman model. First, we analyze their
conformational free energy in several different ways and compare our findings
to two dimensional Flory theory. Then, we study the shape-dependent short-time
diffusivity of the trimer and tetramer chains and calculate the full diffusion
tensor as function of instantaneous shape. By also determining the
shape-averaged translational diffusivity, rotational diffusivity and
flexibility for chains of three to six spheres, we show how these scale as
function of chain length. Overall, we find a good agreement between the
experimental measurement and the simulations, except for translational
diffusivity. In that case, we hypothesize that the difference in surface slip
in the experiments, where the substrate has a finite slip length due to the
hydrogel surface, and simulations, where we use a no-slip boundary condition,
lead to the higher translational diffusivity in the experiments. We hope our work aids the study of diffusivity of
flexible objects found in complex mixtures relevant in, for example, the
cosmetic, pharmaceutical and food industries, as well as in biological systems.
There, our findings may have implications for understanding both the diffusive
behavior and the most likely conformations of macromolecular systems, such as
polymers, single-stranded DNA and other chain-like molecules.

\section{Materials and Methods}

\subsection{Experimental} 

Flexible chains of colloidal supported lipid bilayers (CSLBs) were prepared as
described in previous work
\cite{VanDerMeulen2013,VanDerMeulen2015,Chakraborty2017,Rinaldin2019},
specifically, we followed the exact same procedure as in \cite{Verweij2020} and
used silica particles of two different radii to test the generality of our
findings. We now briefly summarize the experimental procedure from
\citetalias{Verweij2020}.

\begin{figure*} \centering \includegraphics{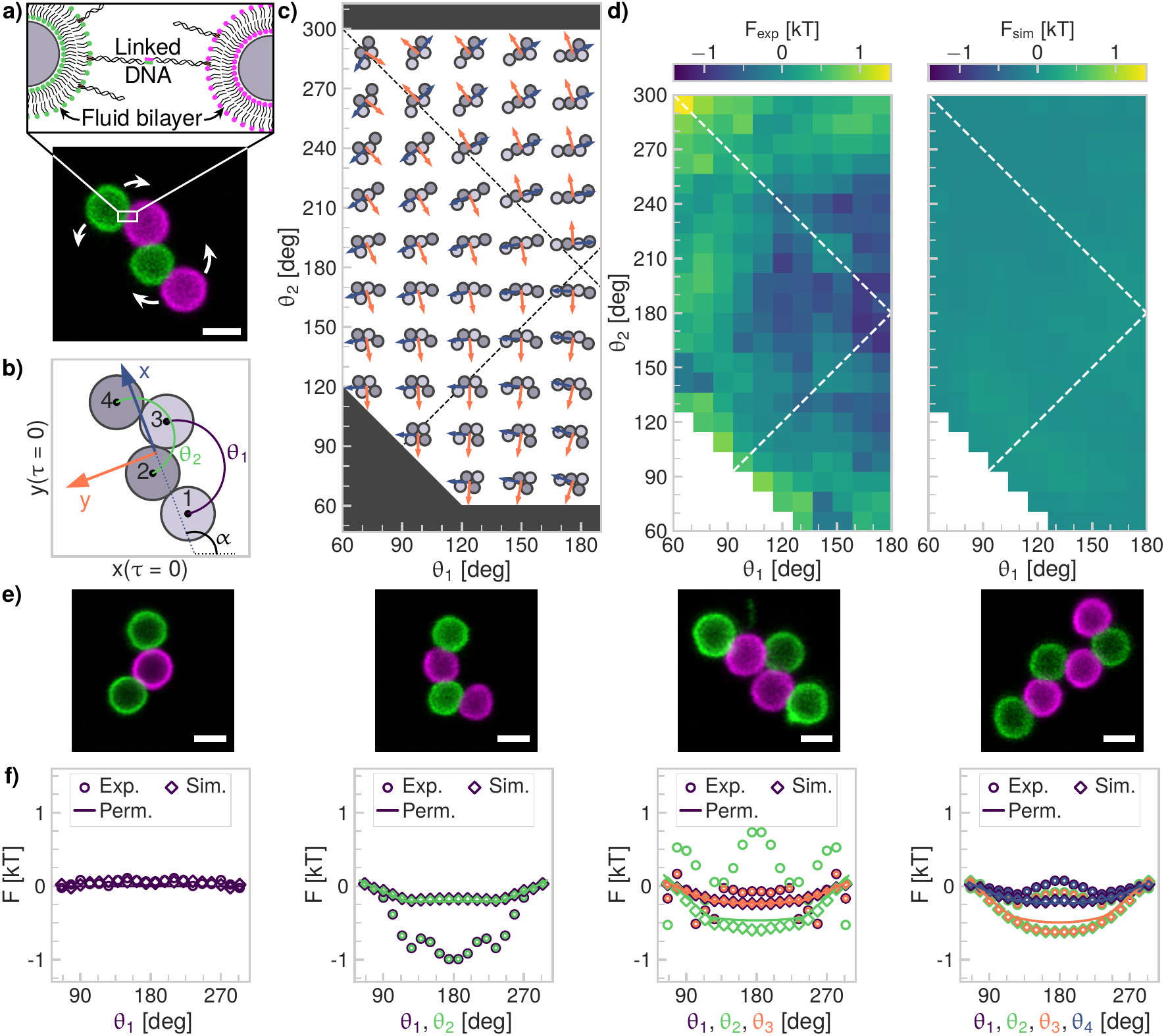}
    \caption{\textbf{Flexibly linked colloidal chains.} \textbf{a)} Flexibly
        linked colloidal chains are created from DNA functionalized colloid
        supported lipid bilayers (CSLBs). The particles are bound by the DNA
        linkers, which can diffuse in the fluid lipid bilayer, yielding
        reconfigurable assemblies. Bottom: Confocal image of a tetramer chain,
        where the different colors, stemming from fluorescently labeled lipids,
        indicate the two different particle types that are functionalized with
        complementary DNA linkers. Scalebar is \SI{2}{\um}. \textbf{b)} The
        coordinate system used for quantifying the diffusion of tetramer
        chains, relative to the center of diffusion (see
        \autoref{sec:trackingpoint}). \textbf{c)} Schematics of possible
        configurations for the tetramer, as function of the two opening angles
        $\theta_{1},\theta_{2}$. Some configurations are sterically prohibited
        because the particles cannot interpenetrate (as indicated by the dark grey
        area). The dashed lines indicate the two symmetry axes of the opening
        angles, $\theta_{2}=\theta_{1}$ and
        $\theta_{2}=\SI{360}{deg}-\theta_{1}$. \textbf{d)} The free energy of
        both experimental (left) and simulated (right) chains of four particles
        in terms of $\theta_{1},\theta_{2}$.  \textbf{e)} Confocal microscopy
        images of, left to right, a trimer ($n=3$), tetramer ($n=4$), pentamer
        ($n=5$) and hexamer ($n=6$) chain.  Scalebars are \SI{2}{\um}.
        \textbf{f)} The free energy in terms of the opening angles of groups of
        three particles (``trimer segments''), of, left to right, trimers,
        tetramers, pentamers and hexamers. The main contribution to the free
    energy is the configurational entropy of the chains. $\circ$ Experimental,
$\diamond$ simulated and permutation data (Perm.) are shown, different colors
indicate different opening angles.\label{fig:fig1}} \end{figure*}

The CSLBs consisting of \SI{2.12\pm0.06}{\um} silica particles were prepared as
described in our recent works \cite{Rinaldin2019,Verweij2020}. Briefly, the
particles were coated with a fluid lipid bilayer by deposition and rupture of
small unilamellar vesicles consisting of \SI{98.8}{\mole\percent} DOPC
(($\Delta$9-Cis) 1,2-di\-o\-le\-oyl-sn-gly\-ce\-ro-3-phos\-pho\-cho\-line),
\SI{1}{\mole\percent} DOPE-PEG(2000)
(1,2-di\-o\-le\-oyl-sn-gly\-ce\-ro-3-phos\-pho\-e\-tha\-nol\-a\-mine-N-[me\-thox\-y(po\-ly\-e\-thy\-lene
gly\-col)-2000]) and \SI{0.2}{\mole\percent} TopFluor-Cholesterol
(3-(di\-pyr\-ro\-me\-thene\-bo\-ron
di\-flu\-o\-ride)-24-nor\-cho\-les\-te\-rol) or DOPE-Rhodamine
(1,2-di\-o\-le\-oyl-sn-gly\-ce\-ro-3-phos\-pho\-e\-tha\-nol\-a\-mine-N-(lis\-sa\-mine\-rho\-da\-mine
B sulf\-o\-nyl)). The bilayer coating was performed in a buffer at pH 7.4
containing \SI{50}{\mM} sodium chloride (NaCl) and \SI{10}{\mM}
4-(2-Hy\-dro\-xy\-e\-thyl)-1-pi\-per\-a\-zine\-e\-thane\-sul\-fo\-nic acid
(HEPES). We added double-stranded DNA (of respectively strands DS-H-A and
DS-H-B, see the Supplementary Information of \citetalias{Verweij2020}) with an 11
base pair long sticky end and a double stearyl anchor, which inserts itself
into the bilayer via hydrophobic interactions (see \autoref{fig:fig1}a, top
panel). The sticky end of strand DS-H-A is complementary to the sticky end of
strand DS-H-B, which allows them to act as linkers. Self-assembly experiments
were performed in a different buffer of pH 7.4, containing \SI{200}{\mM} NaCl
and \SI{10}{\mM} HEPES. Chains of \SI{2.12}{\um} CSLBs were formed by
self-assembly in a sample holder made of po\-ly\-a\-cryl\-a\-mide (PAA) coated
cover glass~\cite{Verweij2020}. Confocal microscopy images of the coated
particles are shown in \autoref{fig:fig1}e, for chain lengths of $n=3$ to 6
particles.

Additionally, we analyzed chains of \SI{1.93\pm0.05}{\um} CSLBs, with silica
particles purchased from Microparticles GmbH (product code SiO$_2-$R-B1072).
We followed the same protocol with two minor modifications: first, the lipid
composition was \SI{91.2}{\mole\percent} DOPC, \SI{8.7}{\mole\percent}
DOPE-PEG(2000) and \SI{0.1}{\mole\percent} DOPE-Fluorescein. Second, we added
double-stranded DNA with a self-complementary 12 base pair sticky end (i.e. a
palindromic sequence) and a cholesterol anchor that inserts itself into the
lipid bilayer due to hydrophobic interactions (see Supplementary Material of
\citetalias{Verweij2020}, strands PA-A and PA-B). To image the \SI{1.93}{\um} CSLBs
we used a flow cell coated with po\-ly\-(2-hy\-dr\-o\-xy\-e\-thyl a\-cry\-late)
(pHEA) polymers~\cite{Verweij2020}. Self-assembly experiments were performed in
a buffer of pH 7.4, containing \SI{50}{\mM} NaCl and \SI{10}{\mM} HEPES.

\subsection{Microscopy}

Chains were imaged for at least \SI{5}{\minute} (frame rates between 5 and
\SI{19}{fps}) at room temperature using an inverted confocal microscope (Nikon
Eclipse Ti-E) equipped with a Nikon A1R confocal scanhead with galvano and
resonant scanning mirrors. A \SI{60}{$\times$} water immersion objective
(NA=1.2) was used. 488 and \SI{561}{\nm} lasers were used to excite,
respectively, the Fluorescein or TopFluor and Rhodamine dyes. Laser emission
passed through a quarter wave plate to avoid polarization of the dyes and the
emitted light was separated by using $500-\SI{550}{\nm}$ and
$565-\SI{625}{\nm}$ filters.

To complement the data obtained from self-assembled chains, we used optical
tweezers to assemble specific chain lengths. Briefly, we employed a homemade
optical setup consisting of a highly focused trapping laser manufactured by
Laser QUANTUM (\SI{1064}{\nm} wavelength). The laser beam entered the confocal
microscope through the fluorescent port, after first passing through a beam
expander and a near-infrared shortpass filter. The same objective was used for
imaging and to focus the trapping laser beam. During the trapping, the quarter
wave plate was removed from the light path.

Particle positions were tracked using a custom algorithm \cite{Rinaldin2019}
available in TrackPy by using the \texttt{locate\_brightfield\_ring} function
\cite{TrackPy} or using a least-square fit of a Mie scattering based model
implemented in HoloPy~\cite{HoloPy}. Both methods agree to an accuracy of at
least \SI{1}{px}, however we have found that the Mie scattering based model is
more robust for tracking multiple particles in close proximity to each other.
For all analysis, we only selected clusters that showed all bond angles during
the measurement time, experienced no drift and were not stuck to the substrate.
An overview of the total number of measurements, the total duration and the
total number of frames per chain length is shown in \autoref{table:data}.

\begin{table}
    \caption{Overview of the amount of measurements, the total duration and the total number of frames per chain length $n$, for the experimental and simulated data.\label{table:data}}
    \begin{tabularx}{\linewidth}{l X S[table-format=3] S[table-format=2] X S[table-format=3] S[table-format=4] X S[table-format=1.1e1] S[table-format=1.1e1]}
        \toprule
        {$n$} && \multicolumn{2}{c}{Measurements} && \multicolumn{2}{c}{Total length [min]} && \multicolumn{2}{c}{Total frames} \\
          && {Exp.} & {Sim.} && {Exp.} & {Sim.} && {Exp.} & {Sim.} \\ \midrule
        3 && 30 & 30 && 275 & 900 && 9.3e4 & 3.8e7 \\
        4 && 69 & 50 && 305 & 1500 && 2.5e5 & 6.4e7 \\
        5 && 13 & 20 && 75 & 600 && 4.7e4 & 2.5e7 \\
        6 && 5 & 20 && 41 & 600 && 4.1e4 & 2.5e7 \\
        \bottomrule
    \end{tabularx}
\end{table}

\subsection{Simulations}\label{sec:simulations}

We have performed Brownian dynamics simulations with hydrodynamic interactions
following the method outlined in \textcite{Sprinkle2020} using the open-source
RigidMultiblobsWall package~\cite{RigidMultiblobsWall}. Hydrodynamic
interactions are calculated using the Stokes equations with no-slip boundary
conditions.  The hydrodynamic mobility matrix is approximated using the
Rotne-Prager-Blake (RPB) tensor~\cite{Swan2007}, which is a modified form of
the Rotne-Prager-Yamakawa (RPY)
tensor~\cite{Rotne1969,Yamakawa1970,Wajnryb2013} and accounts for a bottom
wall, which is unbounded in the transverse directions. These corrections to the
RPY tensor are combined with the overlap corrections described in
\textcite{Wajnryb2013} to prevent particle-particle and particle-wall overlap.
The RPB mobility inaccurately describes near-field hydrodynamic interactions
and therefore breaks down for small separation distances. This can be overcome
by adding a local pairwise lubrication correction to the RPB resistance matrix
as described in detail in \textcite{Sprinkle2020}. Based on the full
lubrication-corrected hydrodynamic mobility matrix, the Ito overdamped Langevin
equation is solved to describe the effect of thermal fluctuations.

\begin{table}
    \caption{\textbf{Permutation data.} For all chain lengths $n$, we generated the $P(N_\theta, n-2)$ configurations obtained by permuting all possible combinations of opening angles. Interpenetrating configurations, which are forbidden due to short-range repulsive forces between particles, were removed from this permutation data and the percentages of these configurations relative to the total number of configurations between 60 and \SI{300}{deg}, as well as between 0 and \SI{360}{deg} (fully freely-jointed case) are shown.\label{table:overlap}}
    \begin{tabularx}{\linewidth}{l X S[table-format=1.2] X S[table-format=1.1e1] X S[table-format=2.1] X S[table-format=2.1]}
        \toprule
    {$n$} && {$\delta\theta$ [deg]} && {$P(N_{\theta},n-2)$} && {\shortstack{Interpen. [\%]\\(60-\SI{300}{deg})}} && {\shortstack{Interpen. [\%]\\(0-\SI{360}{deg})}} \\ \midrule
        3 && 0.01 && 2.4e4 && 0 && 33.3 \\
        4 && 0.04 && 4.4e7 && 6.3 && 58.3 \\
        5 && 0.68 && 4.4e7 && 15.3 && 74.6 \\
        6 && 3.00 && 4.3e7 && 27.1 && 84.6 \\ \bottomrule
    \end{tabularx}
\end{table}

We include a gravitational force on the particles to confine them to diffuse
close to the bottom wall, as in the experiments. Inter-particle bonds are
modeled by harmonic springs of stiffness $1000 k_{B}T/R^2$ and equilibrium
length $2R$, where $R=\SI{1.06}{\um}$ is the particle radius. The bond angle is
not restricted. We set the temperature $T=\SI{298}{\kelvin}$, the viscosity of
the fluid $\eta=\SI{8.9e-4}{\pascal\second}$, the gravitational acceleration
$g=\SI{9.81}{\meter\per\second\squared}$, the particle mass $m_{p} =
\SI{9.5e-15}{\kilo\gram}$ (by assuming a particle density of
\SI{1900}{\kilo\gram\per\cubic\meter}) and the simulation timestep $\Delta t =
\SI{1.42}{\milli\second}$. For the firm potential that prevents overlap, we use
a strength of $4 k_{B}T$ and a cutoff distance $\delta_{\mathrm{cut}} =
\num{e-2}$~\cite{Usabiaga2017,Sprinkle2020}. We initialized the particle chains
in a linear configuration (all opening angles \SI{180}{\degree}). Then, these
initial configurations were randomized by running the integration for a
simulated time of \SI{60}{\second} prior to saving the configurations, to
ensure a proper equilibration of the particle positions, bond lengths,
velocities and opening angles. The particle positions were saved every 8
simulation steps to obtain a final framerate of approximately \SI{90}{fps}.  An
overview of the total number of simulations, the total duration and the total
number of saved frames per chain length is shown in \autoref{table:data}.

For comparison to the simulated and experimental data, we generated data in
which the chains are completely non-interacting and freely-jointed up to steric
exclusions in the following manner: we generated all $(n-2)$-permutations of
the $N_{\theta}$ opening angles $\theta_i$, which gives a total number of
$P(N_{\theta},n-2) = N_{\theta}! / (N_{\theta} - (n-2))!$ combinations of
$\theta_i$. Here, the number of opening angles is
$N_{\theta}=(360-2\times60)/(\delta\theta)$, where $\delta\theta$ denotes the
bin width. Then, we removed those combinations that are forbidden because of
steric exclusions between particles, resulting in the final allowed
combinations, which we call ``permutation data''. In \autoref{table:overlap},
we show the bin widths $\delta\theta$ for each $n$, as well as the total number
of generated permutations $P(N_{\theta},n-2)$. The percentage of permutations
that was removed due to steric exclusions is shown, as well as the total number
of configurations that would result in interpenetrating particles for the
completely freely-jointed case, where $\theta$ can vary between 0 and
\SI{360}{deg}.

\subsection{Data analysis}\label{sec:analysis}

For all fits reported in this work, we used a Bayesian method to find an estimate
of the posterior probability distribution, by using an Affine Invariant Markov
chain Monte Carlo (MCMC) Ensemble sampler method as implemented in the Python
packages \texttt{lmfit}~\cite{lmfit} and \texttt{emcee}~\cite{emcee}. This
allowed us to obtain accurate estimates of the error and the maximum likelihood
estimate (MLE) of the parameters. Parameter values were initialized using a
standard least-square fit, appropriate bounds on the parameter values were
implemented as priors. We estimated the autocorrelation time $\tau_{acor}$ of
the MCMC chain using the built-in methods and ran the analysis for at least
$100\tau_{acor}$ steps, where we discarded the first $2\tau_{acor}$ steps
(corresponding to a burnin phase) and subsequently used every other
$\tau_{acor}/2$ steps (known as thinning). We used 500 independent chains (or
walkers). The reported values correspond to the maximum likelihood estimate of
the resulting MCMC chains, the reported uncertainties correspond to the 16th
and 84th percentiles of the obtained posterior probability distribution.

\subsection{Diffusion tensor analysis}\label{sec:diffusiontensordef}

\subsubsection{Definition of the diffusion tensor}

We determined the short-time diffusivity of the clusters, both as function of
their instantaneous shape, as well as averaged over all possible
configurations. Because the chains are sedimented to the bottom substrate, we
consider only the quasi-2D, in-plane diffusivity. For the flexible trimers, we
followed the methods outlined in \citetalias{Verweij2020}. For the flexible
tetramer chains, we calculated a $5\times5$ diffusion tensor, where the five
degrees of freedom correspond to translational diffusivity in $x$ and $y$,
rotational diffusivity and the flexibilities of the tetramer, which are
described by the diffusivities of the opening angles $\theta_{1},\theta_{2}$.
Specifically, the $x$- and $y$-directions are schematically shown for one
configuration in \autoref{fig:fig1}b and defined by \autoref{eq:xaxis}. The
rotation angle used for determining the rotational diffusivity is indicated in
\autoref{fig:fig1}b and is the angle of the $x(\tau)$ relative to $x(\tau=0)$,
i.e. the angle of the body-centered $x$-axis of the current frame relative to
the body-centered $x$-axis of the reference frame at $\tau=0$. The
flexibilities are calculated from the mean-squared displacements of the opening
angles $\theta_{1},\theta_{2}$, which are depicted in \autoref{fig:fig1}b.
$\theta_{1}$ is defined in such a way that it is always less than or equal to
\SI{180}{deg} and this defines how we assign the magnitude of $\theta_{2}$,
specifically, whether it is acute or obtuse.

The diffusion tensor elements of the tetramer chains were determined
analogously to the trimers~\cite{Verweij2020}. Briefly, for each pair of
frames, we determined the initial shape of the chain, which is determined by
$\theta_{1},\theta_{2}$. The short time diffusion tensor is then calculated
from the trajectories in the following way: \begin{align}
\bm{D}[ij](\theta_{1}, \theta_{2}) &\equiv \frac{1}{2}
\frac{\partial\langle\Delta i (\theta_{1}, \theta_{2}) \Delta j (\theta_{1},
\theta_{2})\rangle_\tau}{\partial \tau}, \label{eq:dtensor} \end{align} with
$\tau$ the lag time between frames, $\langle \cdots \rangle_\tau$ denotes a
time average over all pairs of frames $\tau$ apart and $\Delta i (\theta_{1},
\theta_{2}) = i(\theta_{1}, \theta_{2}, t+\tau) - i(\theta_{1}, \theta_{2}, t)$ is the
displacement of the $i$-th diffusion tensor element, which depends on the
instantaneous shape given by $\theta_{1},\theta_{2}$. The average diffusion
tensor elements $\bm{D}[ij]$ were obtained by fitting the overall slope of the
mean (squared) displacements as a function of lag time $\tau$.  We considered
lag times up to \SI{0.17}{\second}, given by the frame rate of the experimental
data. We only considered trajectories where the variation in
$\theta_{1},\theta_{2}$ did not exceed the edges of the bin describing the
initial shape. Then, we calculated the diffusion tensor elements separately for
each initial shape. For fitting the slopes, we used a MCMC sampling method
described in \autoref{sec:analysis}, where we used a linear model without an
offset.  For longer chains, we only considered the shape-averaged, quasi-2D
translational diffusion coefficient $D_T$, which corresponds to in-plane
diffusivity above the wall. Additionally, we determined the rotational
diffusion coefficient $D[\alpha\alpha]$ from the mean squared angular displacement of the
$x$-axis (defined in \autoref{eq:xaxis}, see \autoref{fig:fig1}b for a
schematic depiction), which describes the rotational diffusivity around an axis
perpendicular to the substrate. Finally, we determine the overall cluster
flexibility $D[\bm{\theta\theta}]$ by calculating the mean squared
displacements of the $(n-2)$ opening angles $\theta_i$ as follows:
\begin{align} \langle \lvert\bm{\Delta\theta}\rvert^2 \rangle &= \langle \lvert
    \left( \Delta\theta_1, \dots, \Delta\theta_{n-2} \right) \rvert^2 \rangle,
    \label{eq:deltatheta}\end{align} so that the flexibility
    $D[\bm{\theta\theta}]$ is given by \begin{align} \langle
        \lvert\bm{\Delta\theta}\rvert^2 \rangle &= 2 (n-2) D[\bm{\theta\theta}]
    t,\label{eq:flexibility} \end{align} analogously to the other diffusion
    tensor elements.

\subsubsection{The influence of the tracking point}\label{sec:trackingpoint}

As tracking point, we considered the center of mass (c.m.) and the center of
diffusion (c.d.), because the choice of origin is expected to affect the
magnitude of the diffusion tensor~\cite{Wegener1985,Cichocki2019}. The c.d. was
calculated from $\bm{A}_{ij}$ defined by Equation 2.16 of \citet{Cichocki2019}
using the RPB tensor~\cite{Swan2007} with lubrication corrections as the
inter-particle mobility matrix $\bm{\mu}_{ij}$.  This tensor includes wall
corrections, as discussed previously in~\autoref{sec:simulations}. The c.d. was
determined from the simulated particle positions, because the height above the
bottom wall was not measured experimentally, but is needed to calculate the
wall corrections. The direction of the body-centered $x$- and $y$-axes was
determined as function of the tracking point $\bm{r}_{t.p.}$, which defines the
origin of the body-centered coordinate frame. We define $\bm{r}_{t.p.} = \rho_1
\bm{r}_1 + \rho_2 \bm{r}_2 + \dots + \rho_n \bm{r}_n$ where $\bm{\rho} =
(\rho_1, \rho_2, \dots, \rho_n)$, which defines the location of the tracking
point as a linear combination of the particle positions (Equation 2.2 and 2.3
of \citet{Cichocki2019}). $\bm{\rho}$ is a weight vector which determines how
much weight is accorded to each particle in the calculation of the tracking
point $\bm{r}_{t.p.}$. As an example, for a trimer, $\bm{\rho} = (1/n = 1/3,
1/3, 1/3)$ when the tracking point is chosen to be the center of mass.

The direction of the $x$-axis was chosen as \begin{align} \bm{\hat{x}} &=
    \pm\left[\dfrac{\bm{r}_{t.p.,1} + \dots + \bm{r}_{t.p.,s_1}}{\rho_{1} +
            \dots + \rho_{s_1}} - \dfrac{\bm{r}_{t.p.,s_2} + \dots +
            \bm{r}_{t.p.,n}}{\rho_{s_2} + \dots +
    \rho_{n}}\right],\label{eq:xaxis} \end{align} where $\bm{r}_{t.p.,i}$ is
    the $i$-th coordinate of the tracking point and the bead chain is split
    into two parts of the same number of particles according to \begin{align}
        \begin{cases} s_1 = s_2 = \lceil \frac{n}{2} \rceil & \text{for odd } n
        \\ s_1 = \lceil \frac{n}{2} \rceil, s_2 = s_1 + 1 & \text{for even } n
    \end{cases} \end{align} Note that for a trimer, with the tracking point at
    the c.m. (i.e. $\bm{\rho} = (1/3, 1/3, 1/3)$), $\bm{\hat{x}}$ is parallel
    to the end-to-end vector, which is the same definition as in our previous
    work~\cite{Verweij2020}.  $\bm{\hat{y}}$ is then chosen such that
    $\bm{\hat{x}}$ and $\bm{\hat{y}}$ form a right-handed coordinate system,
    where the direction of $\bm{\hat{y}}$ is chosen to point away from the
    central part of the cluster towards the tracking point, i.e. along
    $\bm{r}_{t.p.} - \left(\bm{r}_{s_1} + \bm{r}_{s_2}\right)/2$.  This
    orientation was determined for every frame, which fixed the orientation of
    the body-centered coordinate system $\bm{x}(\tau=0), \bm{y}(\tau=0)$. For
    subsequent lag times, the direction of $\bm{y}(\tau)$ was chosen such that
    $\bm{y}(\tau=0)\cdot \bm{y}(\tau) > 0$, i.e. the direction of $\bm{y}$ does
    not change sign. The resulting coordinate system relative to the c.d. is
    visualized for the tetramer chains in \autoref{fig:fig1}b and c.

\section{Results and Discussion}
\subsection{Conformations of flexible chains}\label{sec:res_conformations}

\subsubsection{Shape as function of the trimer segments}

Does a diffusing flexible chain of micron-sized spherical particles have
preferred configurations? This is a natural question to ask, because increasing
the number of spheres per chain increases the percentage of overlapping
configurations (see \autoref{table:overlap}) and could potentially change the
hydrodynamic interactions. We answer this question by considering the free
energy of such chains, which were made by the assembly of colloid supported
lipid bilayers
(CSLBs)~\cite{VanDerMeulen2013,VanDerMeulen2015,Chakraborty2017,Rinaldin2019,Verweij2020}.
These particles are bonded by DNA linkers, which provide specific bonds between
the particles. Because the linkers can diffuse in the fluid lipid bilayer, the
bonded particles can move with respect to each other, as schematically depicted
in \autoref{fig:fig1}a. We compared our experimental data to Brownian dynamics
simulation data, where hydrodynamic interactions between particles and the
substrate are taken into account via the Rotne-Prager-Blake (RPB)
tensor~\cite{Swan2007}, overlap corrections~\cite{Wajnryb2013} and a local
pairwise lubrication correction~\cite{Sprinkle2020} (see
\autoref{sec:simulations} for details). 

We analyzed the free energy of clusters of $n=3$ to 6 particles as function of
their conformation using different methods. For a chain of three particles, a
trimer, a single parameter, the opening angle $\theta$, can describe the
conformations~\cite{Verweij2020}. We have shown before that flexible trimers do
not show a preference for any given opening angle and therefore
conformation~\cite{Rinaldin2019,Verweij2020}. For a chain of four of such
particles (see \autoref{fig:fig1}a for a microscopy image), there are two
angles that characterize the shape of the cluster, $\theta_{1}$ and
$\theta_{2}$, which are the opening angles of the two ``trimer segments''
(groups of three adjacent, bonded spheres) that make up the chain. The
definition of the opening angles is shown in \autoref{fig:fig1}b. For the
tetramer chains, we obtained a 2D-histogram of opening angles for $\theta_{1},
\theta_{2}$ between \SIrange{60}{300}{\degree}, using the simulated and
experimental data. These two internal degrees of freedom lead to a large number
of possible chain configurations, as shown in \autoref{fig:fig1}c.  Some
configurations are forbidden because of steric exclusions, as indicated by the
grey areas. The symmetry lines of the opening angles $\theta_{2}=\theta_{1}$
and $\theta_{2}=\SI{360}{deg} - \theta_{1}$ are indicated as well. The
configurations are symmetric around these lines except for the fact that we
break this symmetry by choosing which angle to label as $\theta_{1}$ and which
as $\theta_{2}$, because this has consequences for the orientation of the body
centered coordinate system, as shown in \autoref{fig:fig1}b and defined in
\autoref{eq:xaxis}.

From the probability density function calculated from the histogram, we
determined the free energy using Boltzmann weighing, \begin{align}
\frac{F}{kT} &= -\ln{p} + \frac{F_0}{kT} \label{eq:boltzmann},
\end{align} where $F$ is the free energy, $k$ is the Boltzmann
constant, $T$ is the temperature, $p$ is the probability density and
$F_0$ is an arbitrary constant offset to the free energy that we have chosen such that the average free energy is equal to zero. Except for steric
restrictions and hydrodynamic interactions, we expect inter-particle
interactions to be weak. Therefore, we hypothesize that there are
mainly entropic contributions to the free energy and that enthalpic
contributions are small. The resulting free energy is shown in
\autoref{fig:fig1}d. Like flexible
trimers~\cite{Rinaldin2019,Verweij2020}, chains of four particles are
freely-jointed, as evidenced by the fact that differences in their free
energy as function of opening angles $\theta_{1},\theta_{2}$ are on the
order of \SI{1}{kT} in \autoref{fig:fig1}d in the experiments and below
\SI{0.1}{kT} in the simulations. Differences smaller than or comparable to the thermal energy are difficult to measure experimentally and are of limited physical relevance. Therefore, we conclude that there is no appreciable preference for any given conformation and the tetramer chains are thus freely-jointed. We use the term ``freely-jointed'' in
the sense that the chains are free to move without any preferred state,
up to steric exclusions stemming from short-range repulsions between
the particles that prevent them from interpenetrating.

\begin{figure*}
    \centering
    \includegraphics{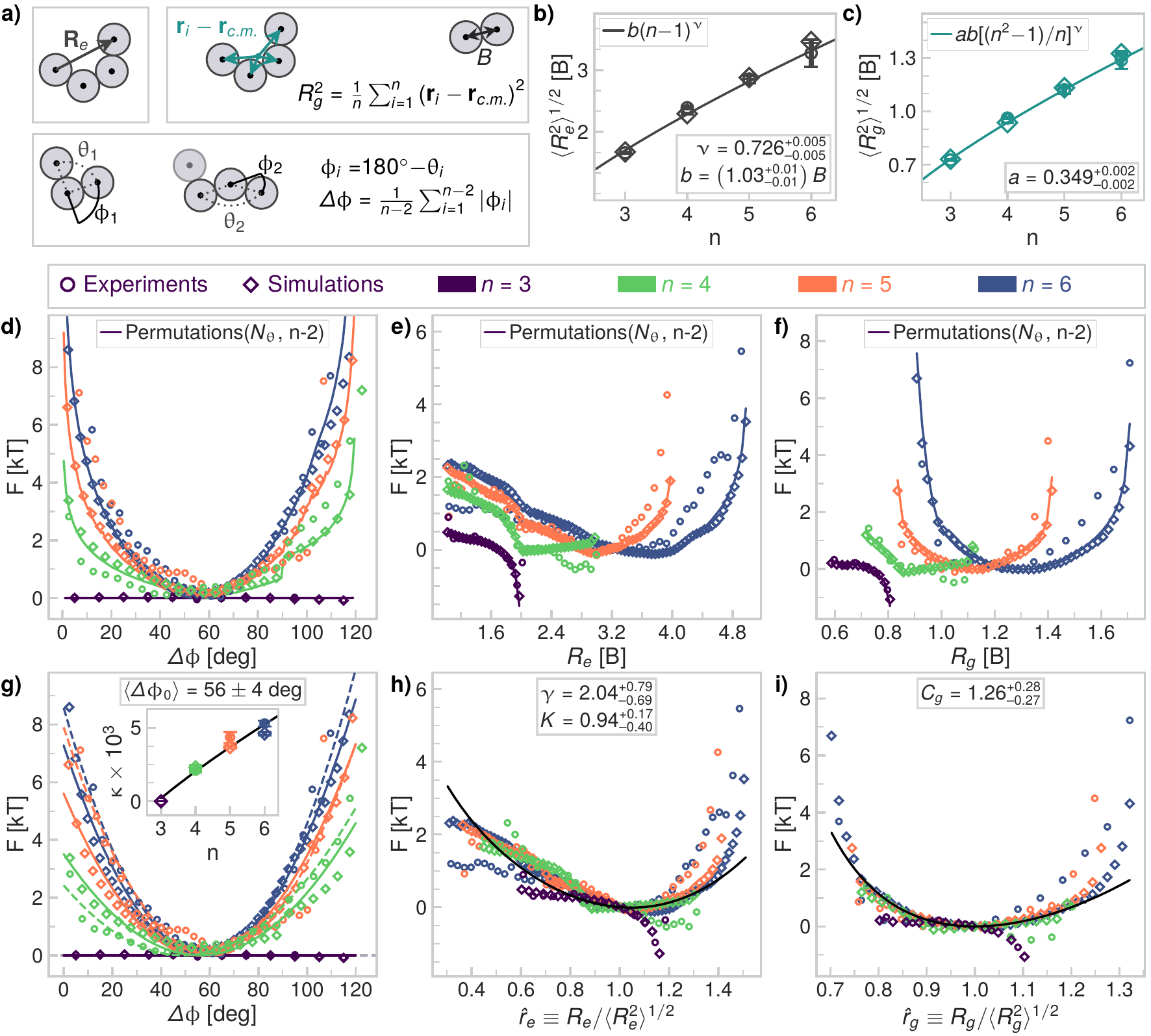}

    \caption{\textbf{Conformations of flexible colloidal chains.} Markers for
        the $\circ$ experimental and $\diamond$ simulated data, together with
        the color coding legend for the different chain lengths $n$, as used in
        panels d-i are shown above those panels. \textbf{a)} Top left: the
        end-to-end distance $R_e$. Top right: definition of the radius of
        gyration $R_g$. Bottom: we define a quantity $\Delta\phi$ which
        measures how much the chain shape deviates from a straight line.
        \textbf{b)} Average $R_e$ and fit of \autoref{eq:avgRe} (fit values
        shown). \textbf{c)} Average $R_g$ and fit of \autoref{eq:avgRg} (fit
        values shown). \textbf{d-f)} Comparison between the free
        energy calculated for experimental, simulated and the
        $P(N_{\theta},n-2)$ permutation data, as function of $\Delta\phi$
        (panel d), $R_e$ (panel e) and $R_g$ (panel f). \textbf{g)} Free
        energy in terms of $\Delta\phi$ and fit of \autoref{eq:harmonic} (fit
        values shown in the insets). Additionally, we fit \autoref{eq:kappa} to
        determine the scaling of $\kappa$ as function of $n$, as shown in the
        inset. \textbf{h)} Free energy in terms of $\hat{r}_e$ and fit of
        \autoref{eq:FRe} (fit values shown).  \textbf{i)} Free energy in terms
of $\hat{r}_g$ and fit of \autoref{eq:FRg} (fit values shown).
\label{fig:fig2}} \end{figure*}

Increasing the number of beads in the chain increases the percentage of
sterically inaccessible configurations (see \autoref{table:overlap}) and could
potentially alter the hydrodynamic interactions. To answer whether these
effects lead to preferred configurations, we study chains consisting of $n$=3
to 6 particles, as shown in confocal microscopy images in \autoref{fig:fig1}e.
Because visualization as a joined histogram becomes increasingly complex as the
chain length increases, we first consider the free energy of their separate
$(n-2)$ opening angles, as shown in \autoref{fig:fig1}f. First, we obtained
probability density functions of their $n-2$ opening angles $\theta_{i}$, where
$\theta_{i}$ is the $i$-th opening angle, as defined in \autoref{fig:fig2}a,
analogously to the choice of opening angles $\theta_{1}, \theta_{2}$ in
\autoref{fig:fig1}b. In other words, we consider the free energy as function of
the opening angles of the trimer segments of the chains. We label the first
particle with $i=1$ and number the rest of the chain consecutively. Because the
first particle on either end of the chain can be given an index $i=1$, there
are also two choices of index labels for each opening angle, except for the
central opening angle when $n$ is uneven. Therefore, we include both choices in
our analysis. Additionally, we include both choices of defining $\theta_1$ as
either obtuse or acute and then use the same convention for the other opening
angles.

Whilst a trimer ($n$=3) shows no preference for any specific
configuration~\cite{Rinaldin2019,Verweij2020}, the tetramer chains do show a
small preference for straight opening angles, as evidenced by the differences
in free energy between compact opening angles and straight opening angles in
\autoref{fig:fig1}f. These deviations are largest in the experimental data, but
also present to a lesser degree in the simulation data and the permutation
data. The deviations in the permutation data can only stem from steric
exclusions, which cause some configurations to be inaccessible: for a tetramer
chain, angles below \SI{60}{deg} or above \SI{300}{deg} and combinations where
$\theta_{1}+\theta_{2}<\SI{180}{deg}$ or $\theta_{1}+\theta_{2}>\SI{540}{deg}$
are not possible due to steric restrictions. Seeing that we have strong
indications that the bond angles are freely-jointed up to steric exclusions in
the simulated and permutation data, we believe that the larger deviations in
the experimental data in \autoref{fig:fig1}d and f, compared to the simulation
and permutation data, are mostly caused by experimental noise. This can be
mitigated by collecting more data, however, the amount of data needed to
characterize the free energy in sufficient angular detail is very large, as can
be inferred from \autoref{table:data}, by comparing the amount of simulated and
experimental data. Because the experimental deviations are below the thermal
energy, we conclude that also in the experiments, there is no preference for
any of the sterically allowed configurations.

For the free energy of the pentamer chains in \autoref{fig:fig1}f, we observe
some larger deviations of the experimental data compared to the permutation and
simulated data. Specifically, the central angle seems to show a preference for
closed angles, as evidenced by the lower free energy for
$\theta_2=\SI{60}{deg}$ and \SI{300}{deg}. However, the difference between the
compact angles and the stretched angles is small, i.e. less than \SI{1.5}{kT}.
Considering the free energy of the hexamer chains, we observe that the
distribution of the outer opening angles $\theta_1,\theta_4$ is flatter than
the distribution of the inner opening angles $\theta_2,\theta_3$, especially in
the simulated and permutation data. Interestingly, there is a clear trend in
the free energy of flexible chains, going from a flat free energy for $n=3$ to
a free energy that shows a minimum at \SI{180}{deg} and becomes increasingly
smooth as the chain length increases. The location of the minimum and the good
agreement with the permutation data show that the deviations from a flat free
energy most likely stem from the steric exclusions at compact opening angles.

\subsubsection{Shape as function of the average bending angle $\Delta\phi$}

So far, we have considered the free energy in terms of the individual opening
angles or trimer segments. To analyze the overall shapes these colloidal chains
can adopt in more detail, we define the average bending angle $\Delta\phi$, as
shown schematically in \autoref{fig:fig2}a.  This allows us to study the
overall shape of the chains by collapsing the $(n-2)$-dimensional description
of the shape in terms of opening angle, onto a single measure of chain shape.
We converted all pairs of opening angles to the single average bending angle,
\begin{align}\Delta\phi = \frac{1}{n-2} \sum^{n-2}_{i=1} \left| \phi_{i}
\right|,\end{align} as defined in \autoref{fig:fig2}a. We then obtained the
probability density function in terms of $\Delta\phi$ and converted this to a
free energy using \autoref{eq:boltzmann}, as shown in \autoref{fig:fig2}d. By
taking this approach, we find that as the number of particles $n$ is increased,
a preferred average bending angle arises at $\Delta\phi_0=\SI{56\pm4}{deg}$,
close to the average of \SI{60}{deg} between no bending (\SI{0}{deg}) and
maximal bending (\SI{120}{deg}), for the experimental, simulated and
permutation data. Additionally, the free energy profiles of all chain lengths
show the same shape, where the difference in free energy between the most
likely and least likely configurations increases as the chain length increases.

The free energy in terms of $\Delta\phi$ effectively quantifies the average
bending angle bending free energy, which is expected to be harmonic based on the
worm like chain model~\cite{wiggins2006generalized}. We fitted a harmonic
potential $V$ to the resulting free energy of the form \begin{align} V =
\frac{\kappa}{2b} \left(\Delta\phi - \Delta\phi_{0}\right)^2
\label{eq:harmonic}, \end{align} with fit parameters $\kappa$ the stiffness of
the potential well in units of $\mathrm{kT/deg^2}$ and $\Delta\phi_{0}$ the
center of the potential well in deg, as shown in \autoref{fig:fig2}g. We find
that the potential well stiffness $\kappa$ increases as the number of particles
increases (see second inset of \autoref{fig:fig2}g) as predicted by polymer
    theory~\cite{wiggins2006generalized}. Namely, we fit \begin{align}
    \kappa&=\alpha \frac{b}{4} (n-1)^{\nu} -\kappa_0, \label{eq:kappa}
\end{align} where $\kappa_0=\SI{5.1\pm0.7e-3}{kT/deg^2}$ fixes the value of
$\kappa$ for $n=3$ and $\alpha=\SI{13\pm1e-3}{kT/deg^2}$ is a positive
constant. We added $\kappa_0$ to the model described by
\textcite{wiggins2006generalized} to ensure that $\kappa=0$ for $n=3$ as we
observe from our data. Additionally, we added the scaling parameter $\alpha$ to
ensure the proper magnitude of $\kappa$. Clearly, for $n=3$, the chain is
freely-jointed and therefore $\kappa=0$. For larger chains, the probability to
observe deviations from a straight configuration decreases as the number of
configurations with steric exclusions increases (see \autoref{table:overlap}),
as is evidenced by the agreement between the permutation, the experimental and
the simulated data in \autoref{fig:fig2}d. Therefore, the bending stiffness
that we measure is an effective parameter, purely arising from these steric
exclusions.

\subsubsection{Distribution of the end-to-end distance and radius of gyration: a comparison with polymer theory}

We now compare the behavior of our flexibly linked particle chains to
predictions from two-dimensional Flory theory for self-avoiding polymers. We
are interested in first testing whether our colloidal chains show the same
behavior as long polymers in terms of RMS end-to-end distance and radius of
gyration, as was shown for chains of flexibly linked oil
droplets~\cite{McMullen2018}. Then, we compare the distributions of the
end-to-end distance and radius of gyration of the colloidal chains to
predictions from polymer theory, to elucidate where finite-size effects start
to play a dominant role in the configurational free energy of chain-like
molecules. In \autoref{fig:fig2}b, c, h and i, we have compared different
measures for the conformations found in our experimental and simulated data,
which we will discuss shortly. We compare these to predictions from polymer
theory by simultaneously fitting \Crefrange{eq:avgRe}{eq:FRg} to the
corresponding values.  We used $a,b,\nu,K,\gamma$ and $C_g$ as fit parameters
and report the maximum likelihood estimates (MLE), the error is given by the
16th and 84th percentiles of the posterior probability distributions, as
explained in detail in \autoref{sec:analysis}.

To compare our data to predictions based on polymer theory, we calculated the
end-to-end distance $R_{e}$, where \begin{align} R_{e} = \left|\bm{r}_{n} -
\bm{r}_{1}\right|, \label{eq:e2e} \end{align} here $\bm{r}_{i}$ is the position
of the $i$-th sphere. Additionally, we determined the radius of gyration
$R_{g}$ as follows, \begin{align} R_{g} = \left[\frac{1}{n^2} \sum^{n}_{i=1}
{\left|\bm{r}_{i} - \bm{r}_{c.m.} \right|^2}\right]^{1/2}, \label{eq:rgyr}
\end{align} with $\bm{r}_{c.m.}$ the center of mass (c.m.) of the cluster. Both
$R_e$ and $R_g$ were normalized by the average interparticle distance $B$ of
each measurement and are schematically shown in \autoref{fig:fig2}a.  For long
polymers, the root mean square (RMS) value of $R_e$ can be described by a power
law~\cite{DeGennes1979},\begin{align}\langle R_e^2 \rangle^{1/2} = b
(n-1)^{\nu},\label{eq:avgRe}\end{align} where $b$ is the Kuhn length (we expect
$b\approx B$) and the Flory exponent ${\nu=3/(d+2)} = 3/4$ for a self-avoiding
walk in $d=2$ dimensions~\cite{Wang1995}. Analogously, the RMS of $R_g$ scales
as~\cite{DeGennes1979} \begin{align}\langle R_g^2 \rangle^{1/2} = a b
[(n^2-1)/n]^{\nu},\label{eq:avgRg}\end{align} where the scaling constant
$a\approx 1/\sqrt{6} \approx 0.41$~\cite{doi1988theory}. We first test whether
the behavior of our colloidal chains is comparable to long polymers in terms of
RMS end-to-end distance and radius of gyration, as was shown for chains of
flexibly linked oil droplets~\cite{McMullen2018}.

Indeed, we find that the scaling of the RMS end-to-end distance and the RMS
radius of gyration of these colloidal chains agree well with the predictions
from polymer theory, as shown in \autoref{fig:fig2}b for the RMS end-to-end
distance and in \autoref{fig:fig2}c for the RMS radius of gyration. For the
Flory exponent we find $\nu=0.726\pm0.005$, which is close to the exact value
of 3/4 for self-avoiding polymers in 2D and in agreement with the value found
for flexibly linked chains of droplets ($\nu=0.72\pm0.03$)~\cite{McMullen2018}.
The Flory exponent is slightly lower than the expected value of 3/4, this might
be explained by the fact that we study a quasi-2D system, in which the
particles have some freedom to move in the out-of-plane direction. This would
lead to a lower value of $\nu$, because in three dimensions
$\nu\approx0.6$~\cite{doi1988theory}. In the simulations, we find an average
center height of $(1.03^{+0.05}_{{-}0.02})R$ above the substrate, calculated
over a random subset of \SI{1}{\percent} of the data, or \num{7e5} positions.
Although these excursions are small, they may lead to the slightly lower value
of $\nu$. On top of that, the slightly lower $\nu$ may be caused by the small
number of beads per chain.

Next, we find that the Kuhn length $b=(\num{1.03\pm0.01}) B$ is in agreement
with the hypothesis that it should be equal to the average bond length. In the
experiments, the bond length is approximately twice the particle radius, plus
the thickness of the bilayer ($\approx\SI{4}{\nm}$) and the length of the DNA
linkers ($\approx\SI{30}{\nm}$). This leads to an estimated experimental bond
length of $B \approx 1.03 (2R)$. In the simulations, because of the harmonic
potential that keeps the particles bonded, we find an average bond length of $B
\approx (\num{1.01\pm0.01}) 2R$, calculated over a random subset of
\SI{1}{\percent} of the data, or \num{6e7} bonds. The fact that the Kuhn length
is slightly greater than the bond length $B$ may be explained by the
greater-than-zero bending stiffness we have found in \autoref{fig:fig2}g. As
shown in \autoref{fig:fig2}c, we find a shape factor $a=0.349\pm0.002$, which
is close to the expected value of $a\approx0.41$~\cite{doi1988theory} and the
value found for flexibly linked chains of droplets
($a=0.30\pm0.02$)~\cite{McMullen2018}.

So far we have found that the RMS end-to-end distance and radius of gyration of
our colloidal chains show the same behavior as long polymers. When we look in
greater detail into the free energy as function of end-to-end distances in
\autoref{fig:fig2}e, we see that our simulated data agrees very well with the
permutation data, as well as the experimental data. Slightly larger deviations
can be seen in the experimental data for $n$=5 and 6, this is due to the fact
that because the number of configurations is very large for longer chains,
increasingly larger amounts of data are needed to probe the equilibrium
distribution, as can be seen from \autoref{table:data} by comparing the amount
of simulated data to the amount of experimental data. 

Based on polymer theory, the free energy of the reduced end-to-end distance
$\hat{r}_{e} \equiv R_{e}/\langle R_{e}^2 \rangle^{1/2}$ should collapse onto a
master curve. For long, self-avoiding polymers, the free energy
$F_e(\hat{r}_e)$ is expected to be equal
    to~\cite{valleau1996distribution,bishop1991investigation} \begin{align}
    \frac{F_e(\hat{r}_e)}{kT} &= (K\hat{r}_{e})^{\delta} - (t+2)
\ln{\hat{r}_{e}} - K^{\delta} \label{eq:FRe} \end{align} with
${\delta=1/(1-\nu)}$, $t=(\gamma-1)/\nu$, $K$ a positive constant and $\gamma$
a positive exponent. This is indeed what we observe in \autoref{fig:fig2}h.
Furthermore, we see the agreement between model and simulated data is better
for longer chains of $n$=5,6 compared to the shorter chains of $n$=3,4, where
finite size effects play a larger role.

Finally, the free energy as function of the radius of gyration in
\autoref{fig:fig2}f is also very well described by the permutation data. It can
be collapsed onto a master curve as function of the reduced radius of gyration
$\hat{r}_{g} \equiv R_{g}/\langle R_{g}^2 \rangle^{1/2}$ as given
by~\cite{Victor1990} \begin{align} \frac{F_g(\hat{r}_g)}{kT} = 2 C_g \left[
    \frac{1}{\alpha}(\hat{r}_{g})^{-d \alpha} +
\frac{d}{\delta}(\hat{r}_{g})^{\delta} + 1 - d \right],\label{eq:FRg}
\end{align} with $\alpha=1/(d\nu-1)$ and $C_g$ a positive normalization
constant. The resulting free energy is shown in \autoref{fig:fig2}i. In terms
of the reduced radius of gyration, deviations from the model are small even for
$n$=4, while only the shortest chains of $n$=3 spheres show some deviations
because of their finite size.

In summary, in this section we have characterized the conformations of flexibly
linked colloidal chains of $n$=3 to 6 spheres. We find that while the chains
are completely freely-jointed, some configurations are forbidden because they
would result in interpenetrating particles. This affects the measured
end-to-end distance and radius of gyration, especially for the shorter chains
of $n$=3 and 4.  Despite these finite size effects, we conclude that the
conformations of all chains can be well described by polymer theory based on
self-avoiding random walks. Based on the generality of the model and the
agreement between model and data, we expect this to be true in general for
other micron-sized objects in which self-avoidance plays a significant role.

\subsection{Shape effects in the diffusion of flexible trimer and tetramer chains}\label{sec:diff_shape}

Having characterized the equilibrium conformations of flexibly linked colloidal
chains, we now analyze their shape-dependent short-time diffusive properties.
Recently, we have studied the effect of flexibility on the diffusivity of the
shortest chain, a freely-jointed trimer~\cite{Verweij2020}. Similar to rigid
particles, we found that shape affects the diffusive motion of the colloid at
short timescales and that displacements are larger in directions that
correspond to smaller hydrodynamic drag. Furthermore, we uncovered a Brownian
quasiscallop mode, where diffusive motion is coupled to Brownian shape changes.
At longer timescales, in addition to the rotational diffusion time, an
analogous conformational diffusion time governs the relaxation of the diffusive
motion, unique to flexible assemblies~\cite{Verweij2020}. 

The choice of coordinate system affects the magnitude of the diffusion tensor.
For all rigid objects, there exist a tracking point relative to which the
diffusion elements are independent of the lag time considered, called the
center of hydrodynamic stress. Although such a point does not exist in general
for flexible objects~\cite{agudo2020diffusion}, an analogous tracking point can
be found where the magnitude of the diffusion tensor elements is minimal and
therefore, close to the time-independent values at long lag times, called the
center of diffusivity (c.d.)~\cite{Cichocki2019}. We compare the results of
two choices of tracking points, namely, the center of mass (c.m.) of the
cluster which is another common choice, and the center of diffusivity
(c.d.)~\cite{Wegener1985,Cichocki2019}. 

The calculation of the c.d. is described in \autoref{sec:diffusiontensordef}.
We find that the c.d. is very close to the c.m. for all chain lengths, but a
slightly larger weight is given to the outer particles compared to the
particles in the center of the chain, as shown in \autoref{fig:com_vs_cod_rho}. The directions of the $x$- and $y$-axis of the
body-centered coordinate system depends on the choice of reference point as
given by \autoref{eq:xaxis}. For the trimer and tetramer chains, using the c.d.
as tracking point, the body centered coordinate systems are shown in
\autoref{fig:fig3}a and \autoref{fig:fig1}b/c, respectively. The rotational
diffusivity is calculated from the angular displacements of the $x$-axis, or
equivalently, rotations around an out-of-plane $z$ axis perpendicular to $x$
and $y$, as indicated in \autoref{fig:fig1}b and \autoref{fig:fig3}a. All
diffusion tensor elements are calculated from \autoref{eq:dtensor}, the method
is explained in detail in \autoref{sec:diffusiontensordef}. The magnitude of
the diffusion tensor elements relative to the c.m. compared to their magnitude
relative to the c.d.\ is given in \autoref{fig:com_vs_cod_trimer} for a
trimer. We find that for a trimer, differences in diffusivities relative to the
c.m. and c.d.\ are only measurable for the rotational-translational coupling
term and the Brownian quasiscallop mode, because the c.m is very close to the
c.d., as is shown in \autoref{fig:fig3}a and \autoref{fig:com_vs_cod_trimer}.

\subsubsection{The diffusivity of flexible trimers: simulations compared to experiments}

\begin{figure*} \centering \includegraphics{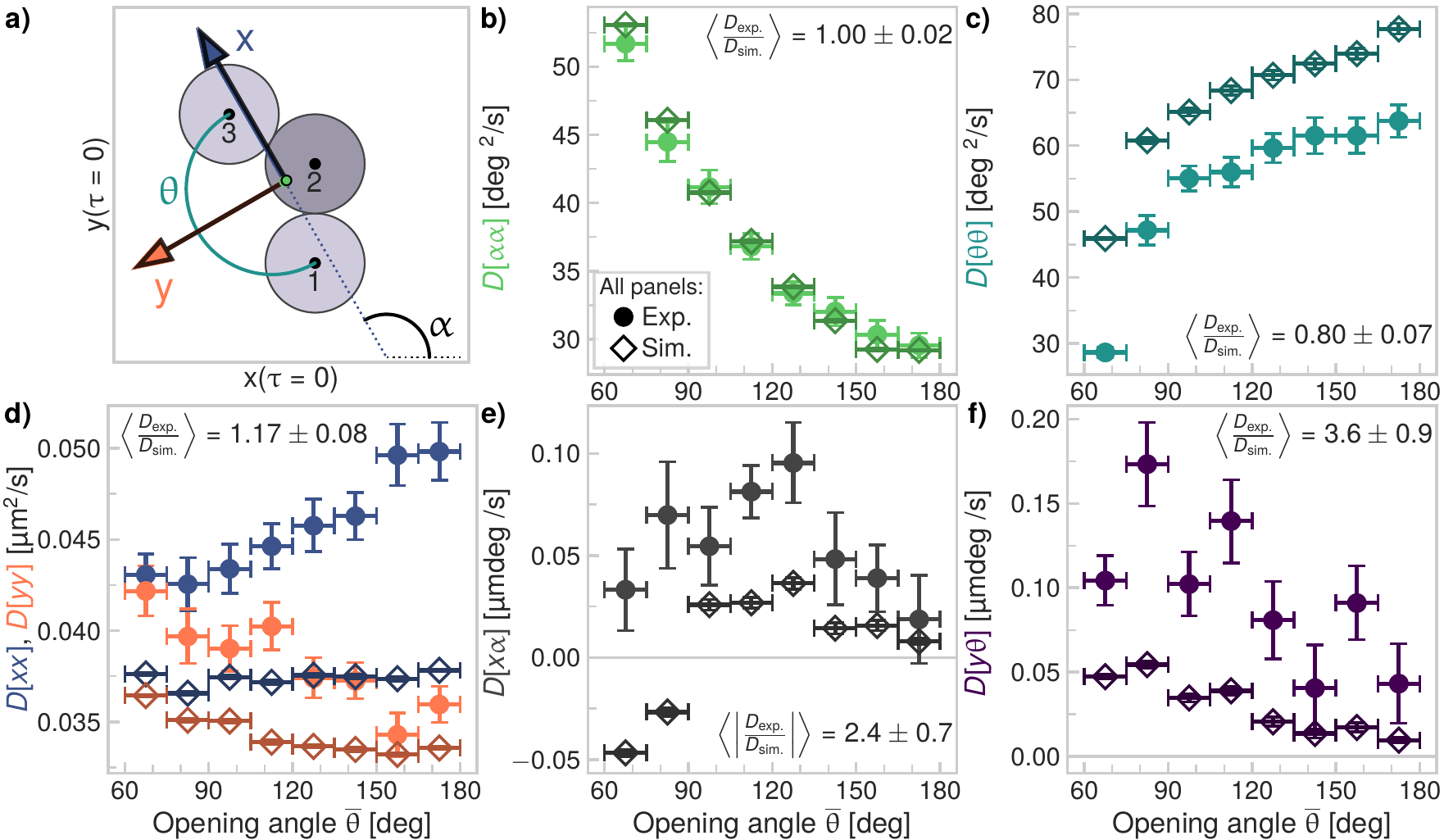}
    \caption{\textbf{Diffusivity of flexible trimers: experimental and
        simulated values.} Comparison between $\bullet$ experimental data and
        $\diamond$ simulated data (both $0.1 \leq \tau \leq
        \SI{0.25}{\second}$). \textbf{a)} Schematic depiction of the coordinate
        system of a trimer, as given by \autoref{eq:xaxis}. The colored arrows
        are calculated with respect to the c.d., while the black arrows are
        relative to the c.m. The difference between the two tracking points
        is very small (see \autoref{sec:SI_cd} for
        details). \textbf{b)} The experimental rotational diffusivity is very
        close to the simulated values.  \textbf{c)} The experimental
        flexibility is lower than the simulated values, because of friction
        stemming from the DNA linker patch.  \textbf{d)} The experimental
        translational diffusivities are larger than the simulated ones, most
        likely caused by the difference in slip conditions in the experiments
        and simulations. \textbf{e)} Translation-rotation coupling term
        $D[x\alpha]$. \textbf{f)} Translation-flexibility coupling term: the
        Brownian quasiscallop mode
$D[y\theta]$~\cite{Verweij2020}.\label{fig:fig3}} \end{figure*}

We have analyzed the diffusivity of flexibly-linked trimers with respect to the
c.d. and we now compare the experimental measurements to our simulations. As
shown in \autoref{fig:fig3}b, for the shape-dependent short-time rotational
diffusivity, there is a quantitative agreement between the experiments and
simulations for most opening angles. Next, we consider the flexibility, which
is defined as half the slope of the mean squared angular displacements of the
opening angle $\theta$ and defines how fast the chain changes its shape,
defined analogously to \autoref{eq:dtensor} for the tetramer chains and given
for the trimers in our previous work~\cite{Verweij2020}. As shown in
\autoref{fig:fig3}c, we measure a lower flexibility in the experiments compared
to the simulations. This is caused by inter-particle friction stemming from the
DNA linker patch embedded in the lipid membrane. Namely, it was found that
increasing the DNA linker concentration leads to a decrease in the
flexibility~\cite{Chakraborty2017}.

Finally, we note that the experimental translational diffusivity is higher than
the translational diffusivity obtained from the simulated data, as can be seen
in \autoref{fig:fig3}d. As a consequence, the experimentally determined
translation-rotation coupling in \autoref{fig:fig3}e, as well as the Brownian
quasiscallop mode in \autoref{fig:fig3}f are also higher than the values
determined from the simulations. This is likely caused by the fact that we
model the substrate as a no-slip surface in the simulations, whereas in the
experiment the substrate is coated with a hydrogel to prevent particles from
sticking, which has a non-zero slip length. We think that this non-zero slip
length in the experiments leads to higher translational diffusivities in the
experiments, because the particles move further away from the glass and so
there is less friction caused by the effectively more viscous water layer close
to the no-slip glass substrate. 

\subsubsection{The diffusivity of flexible colloidal tetramer chains}

\begin{figure*} \centering \includegraphics{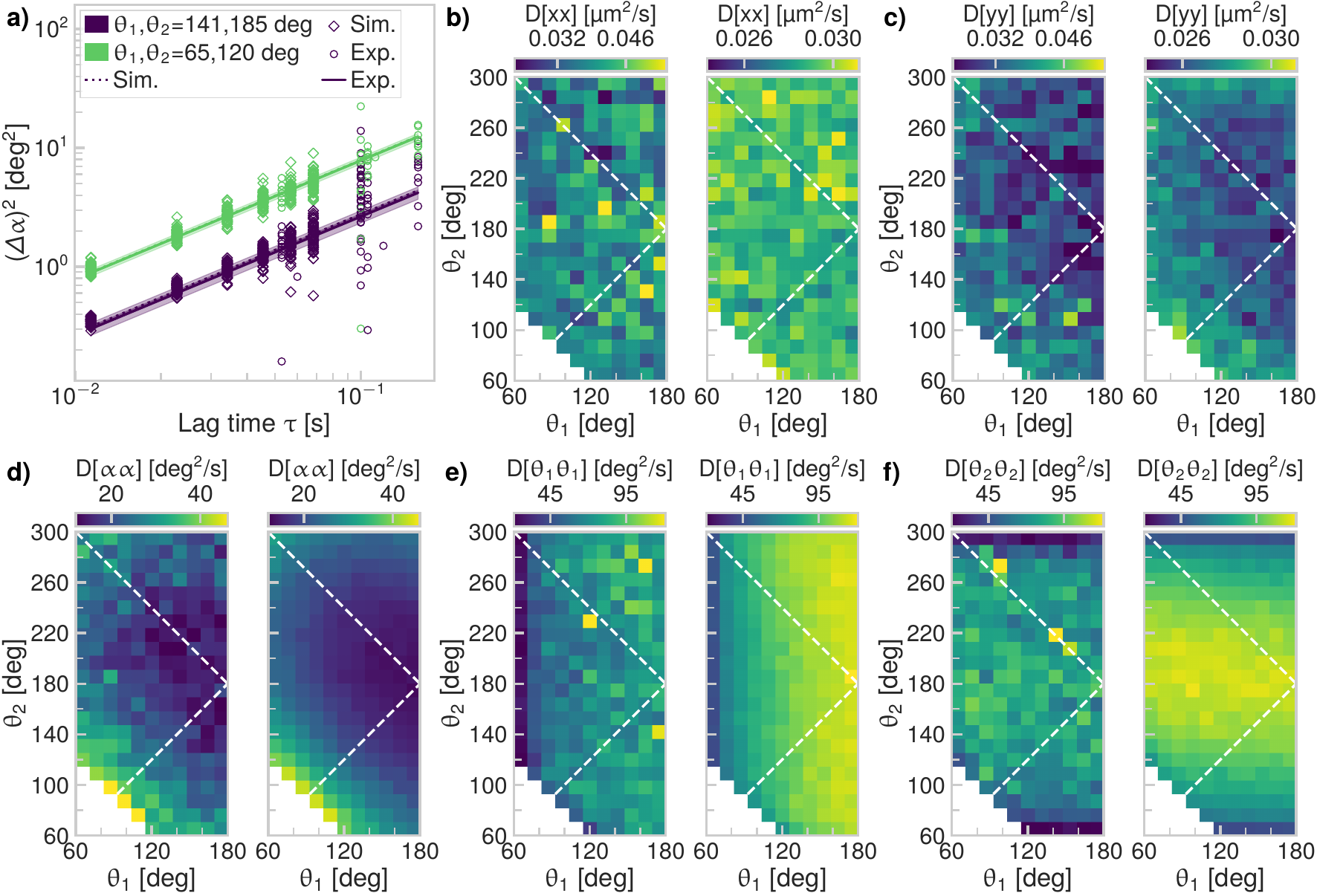}
    \caption{\textbf{Diffusion of flexible tetramer chains.} In panels b-f, the
        left plot shows experimental data and the right plot shows simulated
        data. For fitting, we use a maximum lag time $\tau=\SI{0.17}{\second}$.
        The dashed lines in panels b-f indicate the two symmetry axes of the
        opening angles, $\theta_{2}=\theta_{1}$ and $\theta_{2}=\SI{360}{deg}-\theta_{1}$.
        \textbf{a)} The mean squared angular displacement for a compact
        configuration (green) is larger than that of an extended configuration
        (purple). The elements of the diffusion tensor are obtained by fitting
        the slope of mean squared displacements. \textbf{b)} The translational
        diffusivity in the x-direction as function of the opening angles
        $\theta_{1},\theta_{2}$, as depicted in \autoref{fig:fig1}b and c. \textbf{c)}
        Translational diffusivity in the y-direction, which is lower for chains
        with a stretched angle ($\theta_{1},\theta_{2}$ close to \SI{180}{deg}). For
        both translational terms, we use different color scales for the
        experimental and simulated data, because the experimental diffusivities
        are higher than the simulated ones. \textbf{d)} The rotational
        diffusivity is highest for compact shapes. \textbf{e)} The flexibility
        in the opening angle $\theta_{1}$ shows a maximum for $\theta_{1}=\SI{180}{deg}$.
        \textbf{f)} The flexibility in the opening angle $\theta_{2}$ also has its
    maximum value for $\theta_{2}=\SI{180}{deg}$.\label{fig:fig4}}
    \end{figure*}

Having established that the simulations can faithfully describe the short-time
shape dependent diffusivity of flexible trimers, in addition to the equilibrium
conformations of flexible chains, we now analyze the diffusivity of flexible
tetramer chains. In \autoref{fig:fig4}a, we show that the rotational
diffusivity for compacter shapes is higher than that of more extended shapes
for two examples, $\theta_1,\theta_2=65,\SI{120}{deg}$ (compact) and
$\theta_1,\theta_2=141,\SI{185}{deg}$ (extended). Furthermore, we conclude
that the simulated data agrees with the experimental data, within experimental
error. This can also be seen for the rotational diffusivity as function of
opening angles $\theta_{1},\theta_{2}$ in \autoref{fig:fig4}d. The symmetry
lines of the opening angles $\theta_{2}=\theta_{1}$ and
$\theta_{2}=\SI{360}{deg} - \theta_{1}$ are indicated as well. The
configurations are symmetric around these lines except for the fact that we
break this symmetry by choosing which angle to label as $\theta_{1}$ and which
as $\theta_{2}$, because this has consequences for the orientation of the body
centered coordinate system, as shown in \autoref{fig:fig1}c and defined in
\autoref{eq:xaxis}. However, for the rotational diffusivity we only consider
angular rotations of the $x$-axis and therefore the rotational diffusivity is
indeed symmetric with respect to the symmetry lines of the opening angles.

For the translational diffusivity in the $x$-direction (\autoref{fig:fig4}b)
and $y$-direction (\autoref{fig:fig4}c) we note that the experimental
diffusivity is again slightly larger than the simulated one, similar to the
trimers in \autoref{fig:fig3}d. Again, this is because of the no-slip condition
in the simulations versus the hydrogel surface used in experiments to prevent
particles from sticking, which has a non-zero slip length. Because the
translational diffusivity does not depend on whether the $x$-axis points to one
end of the chain or the other, or equivalently, whether the $y$-axis points
towards one side or the other, we expect that translational diffusivity is
symmetric with respect to the symmetry lines of the opening angles. This is
indeed true: we observe little shape dependence for translational diffusivity
in the $x$-direction in \autoref{fig:fig4}b, variations are likely due to
experimental noise. On the contrary, the diffusivity in the $y$-direction in
\autoref{fig:fig4}c is lower for more extended shapes, which correspond to
larger surface areas and therefore, a larger hydrodynamic drag. We note that it
is also symmetric with respect to the opening angle symmetry lines.

Interestingly, for the diffusivity of the opening angles in \autoref{fig:fig4}e
($\theta_{1}$) and \autoref{fig:fig4}f ($\theta_{2}$), we note that the
flexibility is highest for opening angles $\theta_{1},\theta_{2}$ close to
\SI{180}{deg}, i.e. for the more extended chains. This is in agreement with the
trends we have observed for the flexible trimers in \autoref{fig:fig3}c and our
previous work~\cite{Verweij2020} and suggests that hydrodynamic interactions
between the particles slow down shape changes for small inter-particle
separation distances. Furthermore, we note that the flexibility is not
symmetric around the opening angle symmetry lines, because we have broken the
symmetry in this case, by labeling one angle as $\theta_{1}$ and the other one
as $\theta_{2}$.

The experimental flexibility data in \autoref{fig:fig3}c shows the same trends
as the simulated data but is lower in magnitude. Because the flexibility
depends on the concentration of DNA linkers~\cite{Chakraborty2017}, which cause
additional friction in the bond area, this could also explain the lower
flexibility found in the experimental data, absent in the simulations.
Similarly to what we have found for the trimers in \autoref{fig:fig3}c, we find
that the experimental flexibility in $\theta_{1}$ is \SI{71\pm12}{\percent} of
the simulated one, for $\theta_{2}$ this is \SI{75\pm14}{\percent}. Therefore,
we conclude that the lower magnitude is indeed caused by friction stemming from
the DNA linker patch.

\begin{figure*} \centering \includegraphics{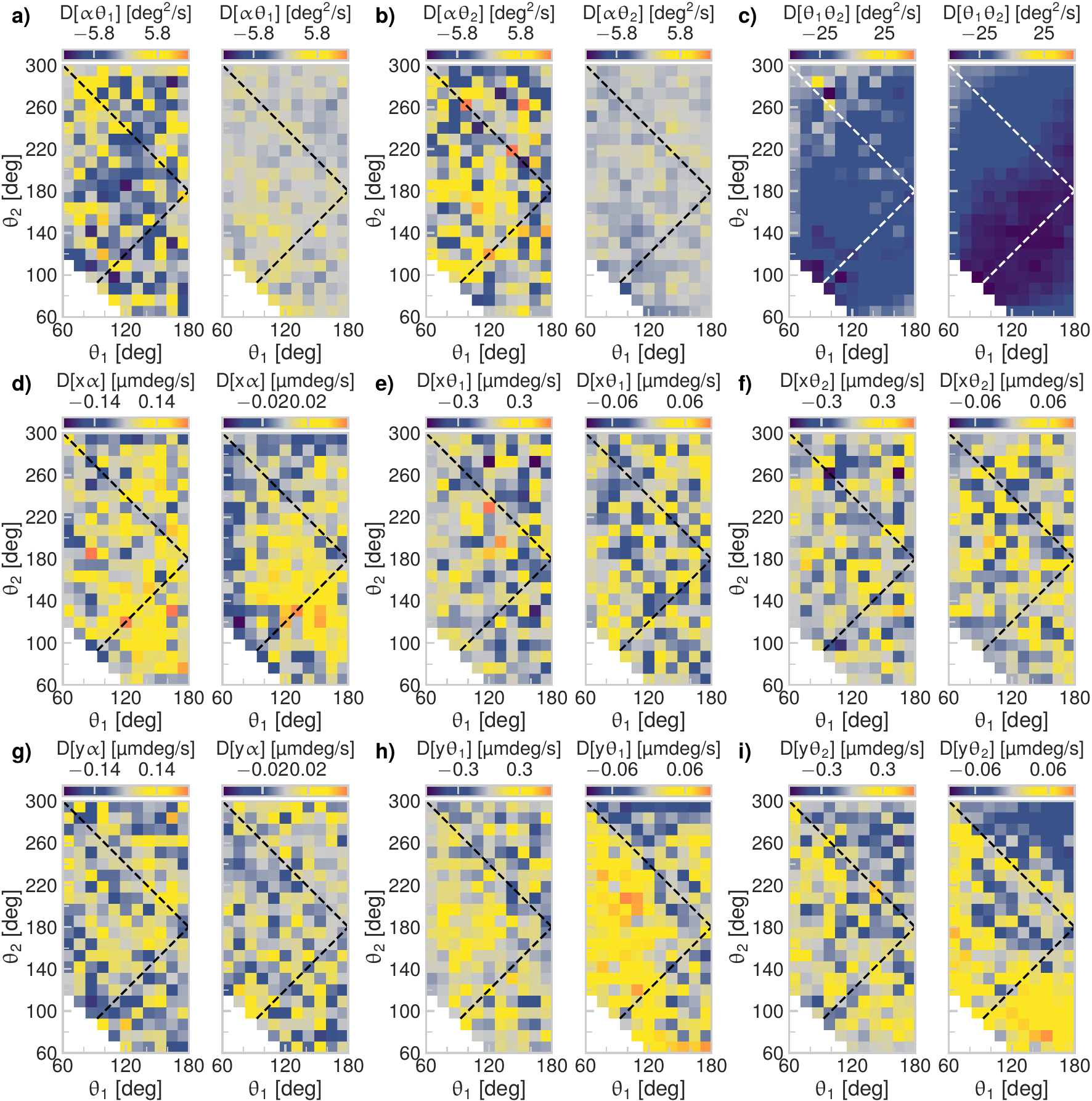}
    \caption{\textbf{Diffusion of flexible tetramer chains: coupling terms.} In
        all panels, the left plot shows experimental data and the right plot
        shows simulated data. The color scales of the experimental and
        simulated data are different for terms involving translational
        diffusivity in panels d-i, because the experimental translational
        diffusivity is higher than the simulated one. For fitting, we use a
        maximum lag time $\tau=\SI{0.17}{\second}$. The dashed lines indicate
        the two symmetry axes of the opening angles, $\theta_{2}=\theta_{1}$ and
        $\theta_{2}=\SI{360}{deg}-\theta_{1}$. \textbf{a)} The rotation-flexibility
        coupling $\mathrm{D[\alpha\theta_{1}]}$ is close to zero. \textbf{b)} The
        rotation-flexibility coupling $\mathrm{D[\alpha\theta_{2}]}$ is also close
        to zero. \textbf{c)} There is a strong negative coupling in
        flexibilities of the two opening angles $\theta_{1},\theta_{2}$. \textbf{d)}
        There is a small translation-rotation coupling $\mathrm{D[x\alpha]}$.
        \textbf{e)} The translation-flexibility coupling $\mathrm{D[x\theta_{1}]}$
        is close to zero. \textbf{f)} The translation-flexibility coupling
        $\mathrm{D[x\theta_{2}]}$ is also close to zero. \textbf{g)} The
        translation-rotation coupling $\mathrm{D[y\alpha]}$ is zero as well.
        \textbf{h)} There exists a non-zero translation-flexibility coupling
        $\mathrm{D[y\theta_{1}]}$.  \textbf{i)} The translation-flexibility coupling
$\mathrm{D[y\theta_{2}]}$ is also non-zero. \label{fig:fig4_coupl}} \end{figure*}

The off-diagonal elements of the diffusion tensor describe possible coupling
terms. We have calculated those terms and find that there is no significant
coupling between rotational diffusivity and flexibility, as shown in
\autoref{fig:fig4_coupl}a for the coupling between $\alpha$ and $\theta_{1}$ and in
\autoref{fig:fig4_coupl}b for $\mathrm{D[\alpha\theta_{2}]}$. This is the same
result we have found for flexible trimers~\cite{Verweij2020} and we hypothesize
this can be generalized to larger chain lengths as well. 

However, there is a strong negative coupling between diffusivities in the two
opening angles $\theta_{1},\theta_{2}$, as shown in \autoref{fig:fig4_coupl}c, which is
symmetric with respect to the symmetry lines of the opening angles. In fact,
the negative coupling is strongest around the symmetry line $\theta_{1}=\theta_{2}$ and
lowest for zig-zag like structures near $\theta_{1}=\SI{60}{deg}$ and
$\theta_{2}=\SI{300}{deg}$. By comparing \autoref{fig:fig4_coupl}c to the
schematics of possible conformations in \autoref{fig:fig1}c, the configurations
where the negative couplings are highest, are those where the outer particles
are both on the same side of the coordinate system, namely the positive $y$
plane. There, the hydrodynamic interactions between the particles are largest
and therefore also the negative coupling is largest.

Interestingly, we also find small, but non-zero coupling terms for
translation-rotation coupling. By comparing the two translation-rotation
coupling terms, we note that $y\alpha$ in \autoref{fig:fig4_coupl}g is small
compared to $x\alpha$ in \autoref{fig:fig4_coupl}d, for both the experimental
and simulated data. This means that displacements in the positive $x$-direction
(see \autoref{fig:fig1}c) will lead to counter-clockwise rotations of the
chain, similarly to the $x\alpha$ coupling we found for the
trimers~\cite{Verweij2020}, as also shown here in \autoref{fig:fig3}e.

Lastly, by comparing the translation-flexibility coupling terms for both
opening angles in the $x$-direction, shown in \autoref{fig:fig4_coupl}e for
$x\theta_{1}$ and \autoref{fig:fig4_coupl}f for $x\theta_{2}$, and in the $y$-direction,
see \autoref{fig:fig4_coupl}h for $y\theta_{1}$ and \autoref{fig:fig4_coupl}i for
$y\theta_{2}$, we observe that the coupling terms related to the $y$-direction are
larger in magnitude than those in the $x$-direction. For the $x$-direction,
there are no clear trends for either opening angle, in both the simulated and
the experimental data, as shown in \autoref{fig:fig4_coupl}e and f. On the
contrary, for translation-flexibility coupling terms in the $y$-direction, we
find couplings, analogously to the Brownian quasiscallop mode we have found for
trimers~\cite{Verweij2020}, also shown in \autoref{fig:fig3}f. 

By looking closely at the coupling between $y$ and $\theta_{1}$ diffusivity in
\autoref{fig:fig4_coupl}h, we observe that the coupling is positive for angles
above the symmetry line $\theta_{1}=\theta_{2}$ and below the symmetry line
$\theta_{1}=\SI{360}{deg}-\theta_{2}$. If we look at the configurations for
these angles in \autoref{fig:fig1}c, we observe that positive $y$-displacements
lead to an opening of one end of the chain, namely the trimer segment with
opening angle $\theta_{1}$, similar to the Brownian quasiscallop mode for
trimers. We see the same effect for configurations below the symmetry line
$\theta_{1}=\theta_{2}$ and in fact, the coupling is symmetric around this
symmetry line. For configurations above the other symmetry line,
$\theta_{1}=\SI{360}{deg}-\theta_{2}$, there are strongly negative correlation
terms, especially near $\theta_{2}=\SI{300}{deg}$. By again studying the
configurations for these angles in \autoref{fig:fig1}c, we note that this is
indeed what would be expected to happen for the trimer segment with opening
angle $\theta_{1}$, based on our earlier findings of the Brownian quasiscallop
mode in trimers. Apart from the expected negative correlations we expect from
the Brownian quasiscallop mode of a trimer, there are also positive values in
this region. However, we cannot compare them directly, because the coordinate
system, and therefore the direction of the $y$-axis, is different in the case
of a tetramer chain. Specifically, it is not centered on the trimer segment.
Therefore, the coupling we have observed is similar to, but more complex than
the Brownian quasiscallop mode in trimers.

Analogously, we observe the same effects for the coupling between diffusivity
in the $y$-direction and the other opening angle $\theta_{2}$, as shown in
\autoref{fig:fig4_coupl}i. Starting below the symmetry line $\theta_{1}=\theta_{2}$, the
coupling is positive, as expected. Above the other symmetry line
$\theta_{1}=\SI{360}{deg}-\theta_{2}$, we observe the opposite, negative coupling, which
is in line with our previous results for the opening angle $\theta_{1}$. Between the
two symmetry lines, something more complicated happens, analogously to the area
above the symmetry line $\theta_{1}=\SI{360}{deg}-\theta_{2}$ for the
$\mathrm{D[y\theta_{1}]}$ coupling discussed previously. There, we observe a
positive coupling for configurations close to $\theta_{1}=\SI{60}{deg}$, as expected
from the Brownian quasiscallop mode and the other coupling term
$\mathrm{D[y\theta_{1}]}$. For the other configurations in the area between the
symmetry lines, we observe both positive and negative coupling terms.
Therefore, we conclude that there the behavior is also more complex than one
would expect based on the assumption that the individual trimer segments show
Brownian quasiscallop modes. This is likely due to the displacement of the
coordinate system from the center of the trimer segment, as well as possible
hydrodynamic couplings between shape changes, as we have observed in
\autoref{fig:fig4_coupl}c.

In summary, in this section we have shown that for both trimers and tetramers,
the flexibility determined from the experimental data is reduced to
approximately 75 to \SI{80}{\percent} of the flexibility determined from
simulations, because of friction of the DNA linker patch, which is not modeled
in the simulations. We have found marked flexibility-induced effects on the
diffusivity of flexible tetramer chains, namely an increase in flexibility for
the more elongated configurations and non-zero couplings between translational
diffusivity and both rotational diffusivity and flexibility, as well as a
strongly negative coupling between diffusivity of the two opening angles. We
have established that the simulations can adequately model our experimental
findings, especially for terms that do not relate to translational diffusivity.
For the translational terms, the slip conditions on the surface play a crucial
role and require further careful consideration in future works.

\subsection{Shape-averaged diffusion of flexible chains}\label{sec:diff_shape_avg}

\begin{figure*} \centering \includegraphics{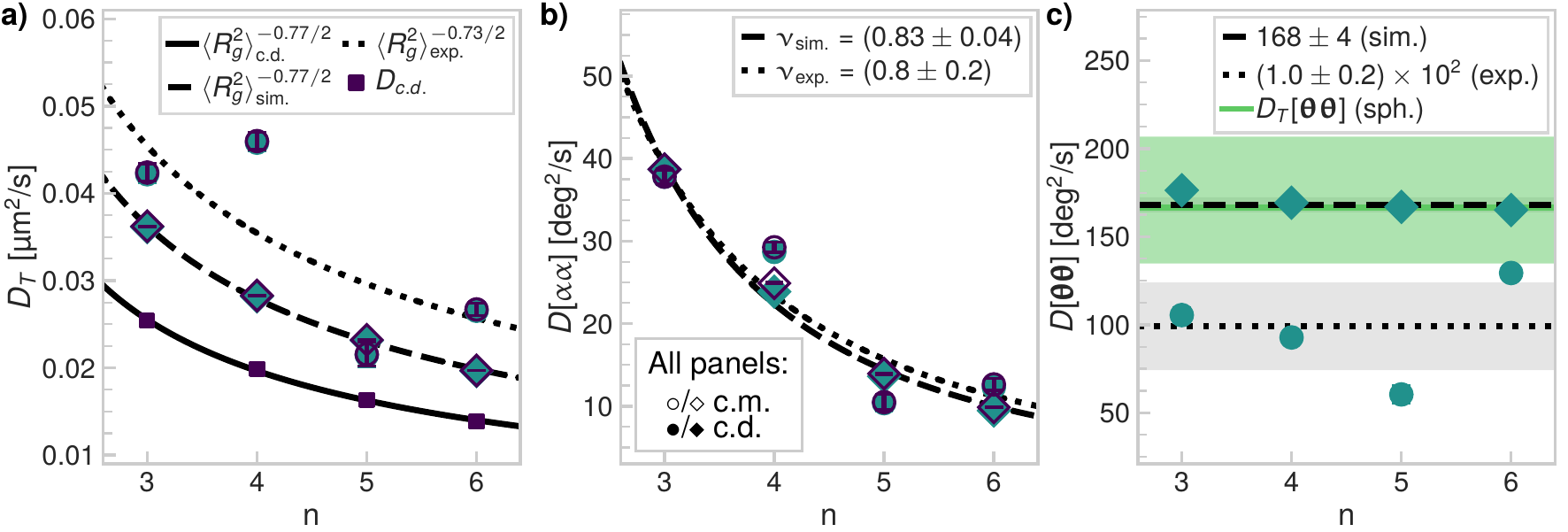}
    \caption{\textbf{Average diffusivity of flexible colloidal chains.} All
        panels: $\circ$ experimental data ($\tau\leq\SI{0.25}{\second}$),
        $\diamond$ simulated data ($\tau\leq\SI{0.05}{\second}$). \textbf{a)} In plane
        translational diffusion coefficient $D_T$ as function of chain length.
        \textbf{b)} Rotational diffusion coefficient $D[\alpha\alpha]$, around the axis
normal to the substrate, as function of chain length. \textbf{c)} Flexibility
coefficient $D[\bm{\theta}\bm{\theta}]$ as function of chain length. The flexibility does not depend
on chain length, the median values plus standard deviations are indicated for
both the experimental and simulated data. \label{fig:fig5}} \end{figure*}

We have studied the short-time, shape dependent diffusivity of flexible trimer
and tetramer chains. For the longer pentamer ($n$=5) and hexamer ($n$=6)
chains, studying the diffusion tensor as function of shape is more challenging
because of the greater number of opening angles and consequently, greater
number of degrees of freedom. Therefore, we take an approach known as the
rigid-body
approximation~\cite{riseman1950intrinsic,Torre1994,Iniesta1988,GarciadelaTorre2002}
and study the shape-averaged short-time diffusivity. 

In \autoref{fig:fig5}a we show the translational diffusivity as function of
chain length for experimental and simulated data, relative to the c.d. and the
c.m. As we have argued before in \autoref{sec:diff_shape}, different choices of
tracking points can lead to different magnitudes of the diffusion tensor
elements. For flexible objects, the c.d. is the most appropriate tracking point
to use, because it gives the smallest values of the diffusion tensor elements
and therefore the obtained values are closer to the long time
diffusivities~\cite{Cichocki2019}. For the flexible colloidal chains, the c.d.\
turns out to be very close to the c.m.\ for all chain lengths, but a slightly
larger weight is given to the outer particles compared to the particles in the
center of the chain, as shown in \autoref{fig:com_vs_cod_rho}.
Because the c.m. is very close to the c.d., we conclude that for our flexibly
linked chains, there is no appreciable difference between the two different
choices of tracking point, both for the experimental and simulated data.

In \autoref{fig:fig5}a, we see a clear scaling of diffusion coefficient $D_T$
with chain length and hypothesize that this scaling can be described by polymer
theory. In Kirkwood-Riseman theory~\cite{riseman1950intrinsic}, the
translational diffusion coefficient is expected to scale as $\langle R_g^2
\rangle^{-\nu/2}$.  Indeed, we find that for the experimental data, the fitted
$\nu=\num{0.7\pm0.5}$ is close to the expected value of $3/4$. More clearly,
for the simulated data, we find $\nu=\num{0.77\pm0.02}$. Again, the average
experimental translational diffusivity is higher than the simulated one,
because of differences in the surface slip conditions, as explained in
\autoref{sec:diff_shape}. Based on the observed behavior, Kirkwood-Riseman
theory can be used to describe the scaling of the translational diffusivity of
the chains as function of their length.

Additionally, we have calculated the lower bound on the short-time diffusion
coefficient $D_{c.d.}$, because its value should be close to the long-time
diffusion coefficient typically measured in scattering
experiments~\cite{Cichocki2019}. We determined $D_{c.d.}$ from the matrix
$\bm{A}_{ij}$ defined by Equation 2.16 of \citet{Cichocki2019} using the RPB
tensor~\cite{Swan2007} with lubrication corrections as the inter-particle
mobility matrix $\bm{\mu}_{ij}$, as explained in
\autoref{sec:diffusiontensordef}. Indeed, we find that its value is always
lower than the simulated or experimental values, which indicates that memory
effects, or in other words, time correlations, play a large role in the
translational diffusivity of these flexible colloidal chains. We find the same
scaling as function of chain length as for the experimental and simulated data,
namely, $\nu=\num{0.77\pm0.02}$, as predicted by Kirkwood-Riseman theory.

Next, having characterized the translational diffusivity of the flexible
chains, we now consider their rotational diffusivity $D[\alpha\alpha]$. While
there is no unique choice for which axis to use to quantify the rotational
diffusivity of a shape-changing object, we use the definition in
\autoref{eq:xaxis} for consistency. As shown in \autoref{fig:fig5}b, the
simulated data agree with the experimental data and the differences between
using the c.m. or the c.d. as tracking point are minimal. We use an approximate
expression to describe the rotational diffusivity of our flexible chains in
the rigid rod limit~\cite{riseman1950intrinsic}: \begin{align} D[\alpha\alpha] &\propto
\frac{\ln\left(2L/d\right)}{L^3}, \end{align} with $L$ the length of the
rod and $d$ its diameter. Setting $d=b$ (the Kuhn length) and
$L=b\left(1+(n-1)^\nu\right)$, which is the average end-to-end distance
plus the Kuhn length, we obtain a reasonable fit with $\nu$ close to the
expected $3/4$, as shown in \autoref{fig:fig5}b. Specifically, we find
$\nu=\num{0.8\pm0.2}$ for the experimental data and $\nu=\num{0.83\pm0.04}$
for the simulated data. Therefore, we conclude that while the shape
dependent short time diffusivity of flexible colloidal chains shows clear
flexibility effects as discussed in \autoref{sec:diff_shape}, the scaling
of the shape-averaged translational and rotational diffusion coefficients
can be described very well by the rigid body approximation.

In \autoref{sec:diff_shape}, we have found couplings between translational
diffusivity and both rotational diffusivity and the flexibility of trimers and
tetramers, for the shape-dependent diffusion tensor. As is shown in \autoref{fig:shape_avg_coupling}, the translation-flexibility coupling term
is averaged out for chains longer than a trimer, presumably because of the
negative correlations between opening angles we have found for the tetramer
chains, which could be present for longer chains as well. The
translational-rotational coupling mode is nonzero for all chain lengths and
decreases as function of chain length, as shown in \autoref{fig:shape_avg_coupling}.

Finally, we consider the shape-averaged flexibility $D[\bm{\theta\theta}]$ by
calculating the slope of the mean squared angular displacements of the $n-2$
opening angles $\langle \lvert\bm{\Delta\theta}\rvert^2 \rangle$ as function of
lag time, as defined by \Crefrange{eq:deltatheta}{eq:flexibility}. First, we
observe that the mean squared displacements of $\bm{\Delta\theta}$ increase
linearly with lag time, similarly to the other diffusion tensor terms.
Moreover, we find that the flexibility is independent of the length of the
chain, as shown in \autoref{fig:fig5}c. 

As we have seen for the shape-dependent short-time flexibilities of the trimer
and tetramer in \autoref{fig:fig3}c and \autoref{fig:fig4}e/f respectively, the
average experimental flexibility is approximately 75 to \SI{80}{\percent} of
the average flexibility in the simulations. We hypothesize that the lower
flexibility in the experiments is caused by inter-particle friction stemming
from the DNA linkers. This is also supported by the observation that while the
experimental flexibility shows large fluctuations, that are due to experimental
differences in DNA concentration between particles and samples, the flexibility
in the simulations shows a very narrow distribution. For the shape-averaged
flexibility, we find that the experimental flexibility averaged over all chain
lengths is \SI{60\pm15}{\percent} of the simulated one, where the spread is
most likely caused by the spread in the DNA linker concentration.

For trimers of flexibly-linked emulsion droplets, it was proposed that for
small displacements of the opening angle, the maximum value of the flexibility
is expected to be dominated by the translational friction coefficient of the
particles~\cite{McMullen2018}. However, in the droplet-based system, the
flexibility was found to be unaffected by the DNA linker
concentration~\cite{McMullen2018}. Because of this difference, we test whether
the same behavior applies to our flexibly linked chains of colloidal particles,
in spite of the presumably different dynamics, caused by the friction of the
DNA linker patch in the lipid bilayer and with the particle surface. By
considering small displacements of the particles, one can define an equivalent
``translational'' flexibility coefficient $D_T[\bm{\theta\theta}]$, which is linked to $D_T$
as $D_T[\bm{\theta\theta}] = (\pi R/\SI{180}{deg})^{-2} D_{T}$. If the flexibility is
dominated by translational diffusivity of the individual spheres, we can
calculate its maximum magnitude from $D_T$. This would in turn mean that the
flexibility scales with the particle radius as $1/R$, in the same way as
translational diffusivity and not as $1/R^3$, as we~\cite{Verweij2020} and
others assumed in previous works~\cite{Harvey1983}.

Indeed, using the methods explained in \autoref{sec:simulations}, for an
individual sphere at a height of $(1.03^{+0.05}_{{-}0.02})R$ above the
substrate, which corresponds to the heights of the spheres in the chains as
found in the simulations, we find a translational diffusion coefficient between
0.046 and \SI{0.071}{\um\squared\per\second}. For comparison, the bulk
diffusion coefficient of the spheres, far from the substrate, is
\SI{0.229}{\um\squared\per\second}. Note that even a small change in the height
above the substrate has a large effect on the calculated diffusion coefficient,
therefore, the spread in the translational diffusion coefficient in
\autoref{fig:fig5} is most likely larger than the reported spread, which is
estimated from the fit of a linear model to the MSD using the method described
in \autoref{sec:analysis}. In the same way, the uncertainty in the expected
flexibility calculated from the translational diffusivity of an individual
sphere is also large. The equivalent range of flexibilities
$D_T[\bm{\theta\theta}]$ based on these values is shown in green in
\autoref{fig:fig5}c. Because the value of the equivalent
$D_T[\bm{\theta\theta}]$ of an individual sphere is very close to the
flexibilities we find in our simulations, we conclude that the maximum
flexibility is indeed set by the translational diffusion coefficient of the
individual spheres. 

\section{Conclusions}

In conclusion, we have studied flexibly linked colloidal chains of three to six
spheres using both simulations and experiments. We have analyzed their
conformational free energy in several different ways. First, we found that the
chains are freely-jointed, except for configurations that are forbidden because
of steric restrictions due to interpenetrating particles. Furthermore, apart
from some deviations because of their finite length, two dimensional Flory
theory for infinitely long polymers can describe their conformational free
energy in terms of reduced end-to-end distance and radius of gyration very
well. We found that the effective bending stiffness, which measures deviations
from opening angles close to straight angles, scales according to the worm-like
chain model.

Then, we have studied the shape-dependent short-time diffusivity of the trimer
and tetramer chains. We found that the simulations can adequately model the
experimental diffusion tensor of flexible trimers. For the flexibly-linked
tetramers, we have found that shape affects the short-time diffusion tensor in
ways similar to what we have found for the shorter trimers.  Namely,
translational and rotational diffusivity are highest in directions that
correspond to the lowest projected surface area, in other words, the more
compact shapes, and the flexibility is highest for the more elongated shapes.
Furthermore, there are non-zero couplings between translational diffusivity and
both rotational diffusivity and flexibility, similar to what we found for the
flexible trimers. Additionally, there is a strong negative coupling between the
diffusivities of the two opening angles. 

By determining the shape-averaged translational and rotational diffusivity for
chains of three to six spheres, we found that these scale as function of chain
length according to Kirkwood-Riseman theory. Their maximum flexibility does not
depend on the length of the chain, but is determined by the near-wall in-plane
translational diffusion coefficient of an individual sphere. The experimental
flexibility is approximately 75 to \SI{80}{\percent} of the flexibility
calculated from the simulated data, because of friction of the DNA linker
patch.

Overall, we found a good agreement between the experimental measurements and
the simulations, except for translational diffusivity. In that case, we
hypothesize that the difference in surface slip in the experiments, where there
is a finite slip length due to the hydrogel surface, and in the simulations,
where we use a no-slip boundary condition, lead to a higher translational
diffusivity in the experiments. We hope our work aids the study of diffusivity
of flexible objects found in complex mixtures relevant in, for example, the
cosmetic, pharmaceutical and food industries, as well as in biological systems.
There, our findings may have implications for understanding both the diffusive
behavior and the most likely conformations of macromolecular systems, such as
polymers, single-stranded DNA and other chain-like molecules.

\subsubsection*{Acknowledgments} We thank Ali Azadbakht for the design and
setup of the Optical Tweezers and his technical support. We are grateful to
Aleksandar Donev and Brennan Sprinkle for fruitful discussions and for
providing us with example code for the simulations. We thank Piotr Szymczak for
useful discussions about the center of diffusion. The simulations were partly
performed using the ALICE compute resources provided by Leiden University.
J.G. wishes to thank the program for Chang Jiang Scholars and Innovative
Research Teams in Universities (no. IRT 17R40) and the 111 Project of the PRC.
This project has received funding from the European Research Council (ERC)
under the European Union's Horizon 2020 research and innovation program (grant
agreement no. 758383) and from the NWO graduate program.

\subsubsection*{Author contributions} RWV, PM, LH, NL and IC performed the
experiments. RWV and LH did the particle tracking, RWV carried out the
simulations and the data analysis. RWV and DJK conceived the experiments and
wrote the paper. All authors discussed the final manuscript.

\appendix\section{Supplementary Movies}

For all movies, a video without annotation is shown on the left and a video
with the annotated particle positions and position of the c.m. in time is shown
on the right. All videos have been sped up five times (original duration
$\SI{5}{\min}$), the pixel size is indicated by a scalebar.

\begin{description}
    \item[\url{MOV_S1_trimer.mp4}] A flexible trimer (n=3).
    \item[\url{MOV_S2_tetramer.mp4}] A flexible tetramer (n=4). 
    \item[\url{MOV_S3_pentamer.mp4}] A flexible pentamer (n=5).
    \item[\url{MOV_S4_hexamer.mp4}] A flexible hexamer (n=6).
\end{description}

\section{Effect of tracking point, bin width and lag time on the diffusivity of flexible trimers}\label{sec:SI_cd}

\begin{figure*}
    \centering
    \includegraphics[width=\linewidth]{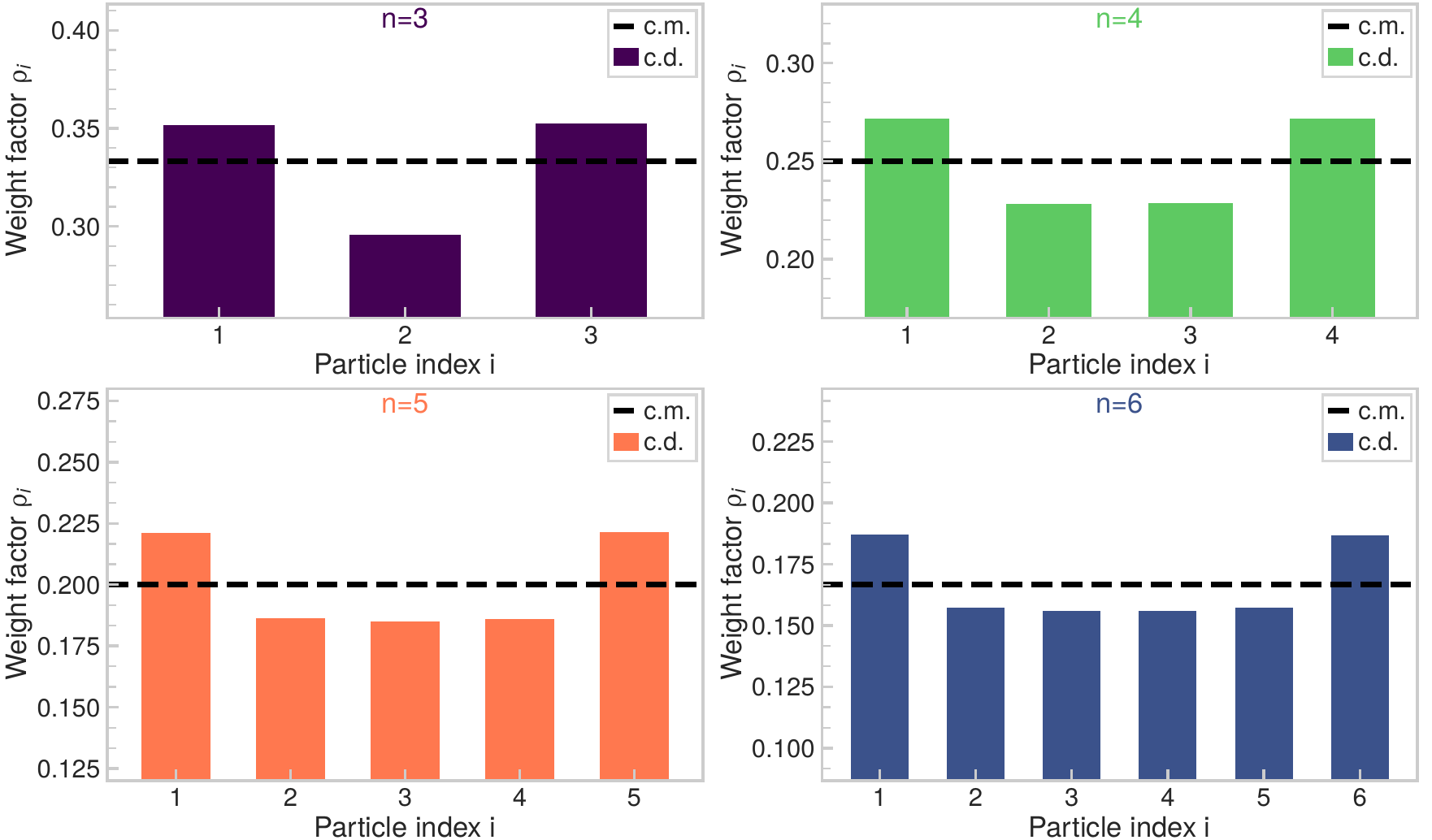}
    \caption[C.d. versus c.m.: weight factors]{\textbf{Center of diffusion (c.d.) versus center of mass (c.m.): weight factors.} Weight factors $\bm{\rho}$ for the c.m. and the c.d., for \textbf{a)} trimer (n=3) \textbf{b)} tetramer (n=4) \textbf{c)} pentamer (n=5) \textbf{d)} hexamer (n=6) chains. For the c.d., more weight is accorded to the outer particles compared to the inner particles of the chain. However, differences between the c.m. and the c.d. are small for all chain lengths.\label{fig:com_vs_cod_rho}}
\end{figure*}

The choice of coordinate system affects the magnitude of the diffusion tensor.
For all rigid objects, there exist a tracking point relative to which the
diffusion elements are independent of the lag time considered, called the
center of hydrodynamic stress. Although such a point doesn't exist in general
for flexible objects~\cite{agudo2020diffusion}, an analogous tracking point can
be found where the magnitude of the diffusion tensor elements is minimal and
therefore, close to the time-independent values at long lag times, called the
center of diffusivity (c.d.)~\cite{Cichocki2019}. We compare the results of
two choices of tracking points, namely, the center of mass (c.m.) of the
cluster which is another common choice, and the center of diffusivity
(c.d.)~\cite{Wegener1985,Cichocki2019}. 

\begin{figure*}
    \centering
    \includegraphics[width=\linewidth]{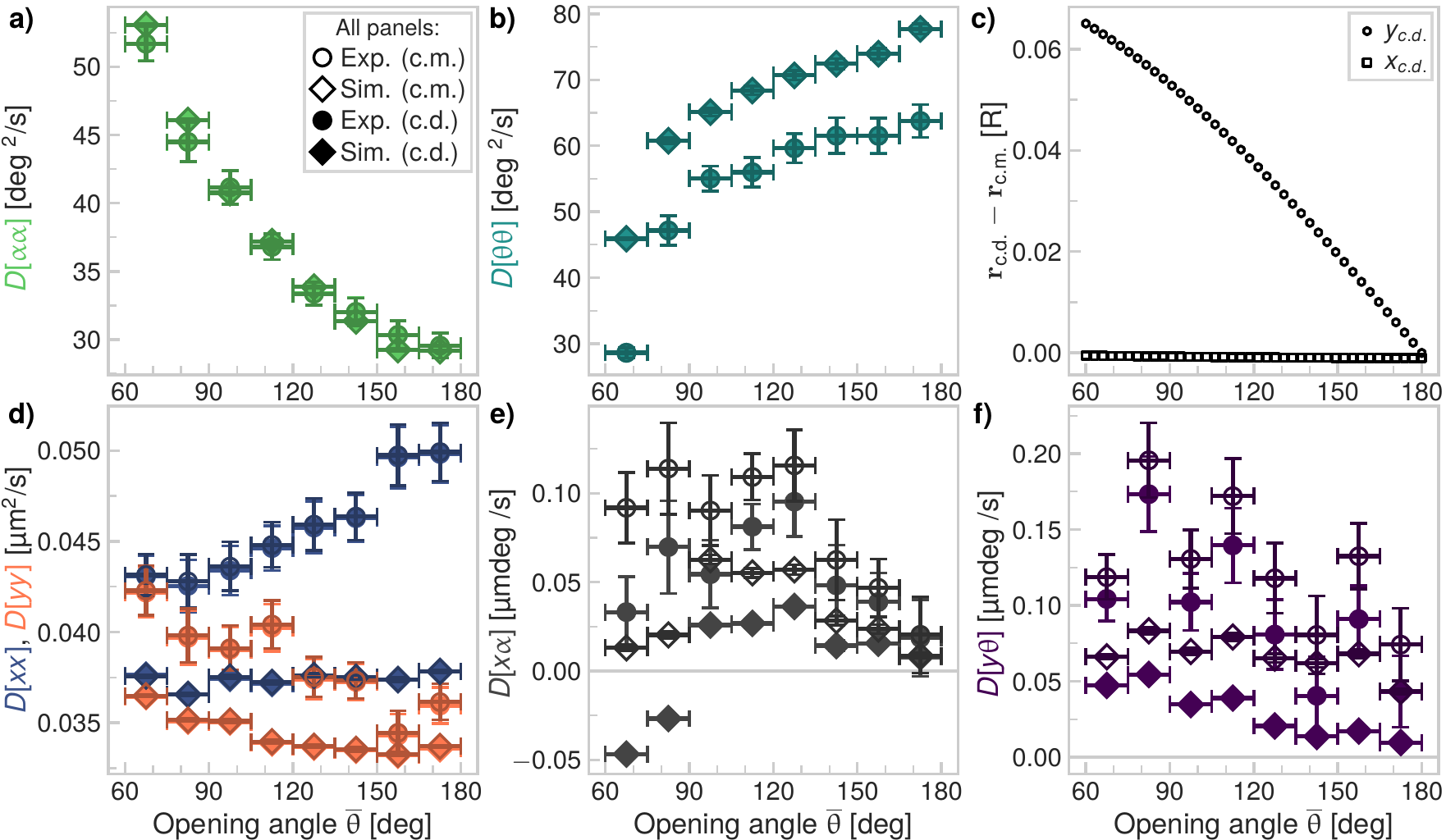}
    \caption[Trimer: c.d. versus c.m. as tracking point]{\textbf{Diffusivity of flexible trimers: center of diffusion (c.d.) versus center of mass (c.m.) as tracking point.} For all panels, open points correspond to the c.m. as tracking point while filled points refer to the c.d. as tracking point. $\circ$ experimental data, $\diamond$ simulated data (both $0.1 \leq \tau \leq \SI{0.25}{\second}$). \textbf{a)} The rotational diffusivity does not change as function of tracking point. \textbf{b)} Also for the flexibility term, there is no influence of tracking point. \textbf{c)} The c.d. of flexible trimers is very close to the c.m.: there is only a small deviation of approximately \SI{6}{\percent} for the smallest opening angles. The difference $\bm{r}_{\mathrm{c.d.}}-\bm{r}_{\mathrm{c.m.}}$ on the $y$-axis is given in terms of the particle radius $R$. \textbf{d)} The translational diffusivity changes only slightly with respect to a different tracking point. \textbf{e)} The effect of tracking point for rotation-translation coupling is more pronounced: values are lower when the c.d. is used as tracking point. \textbf{f)} Also for the Brownian quasiscallop mode, the values are lower when using the c.d. as tracking point instead of the c.m. \label{fig:com_vs_cod_trimer}}
\end{figure*}

First, as shown in \autoref{fig:com_vs_cod_rho}, the c.d. is very close to the
c.m. for all chain lengths. A slightly larger weight is given to the outer
particles for the c.d. compared to the c.m., but this has only a very small
effect on the location of the c.d. Second, we compare the magnitude of the
diffusion tensor of a flexible trimer relative to the c.m. as in our previous
work~\cite{Verweij2020} and relative to the c.d. (this work). As shown in
\autoref{fig:com_vs_cod_trimer}, the rotational
(\autoref{fig:com_vs_cod_trimer}a), flexibility
(\autoref{fig:com_vs_cod_trimer}c) and translational
(\autoref{fig:com_vs_cod_trimer}d) terms of the diffusion tensor are only
slightly affected by changing the tracking point from the c.m. to the c.d. This
is easily explained by the fact that the position of the c.d. only changes by
approximately \SI{6}{\percent} for the smallest opening angle compared to the
c.m., as shown in \autoref{fig:com_vs_cod_trimer}c. However, the coupling terms
are lower with respect to the c.d. as shown in \autoref{fig:com_vs_cod_trimer}e
for the translation-rotation coupling term and in
\autoref{fig:com_vs_cod_trimer}f for the Brownian quasiscallop mode. The fact
that these coupling terms are lower is expected, because the magnitude of the
diffusion tensor is expected to be the lowest relative to the c.d., as it is
closest to the long-time diffusion tensor, for which short-time correlations or
memory effects are expected to vanish.

\begin{figure*}
    \centering
    \includegraphics[width=\linewidth]{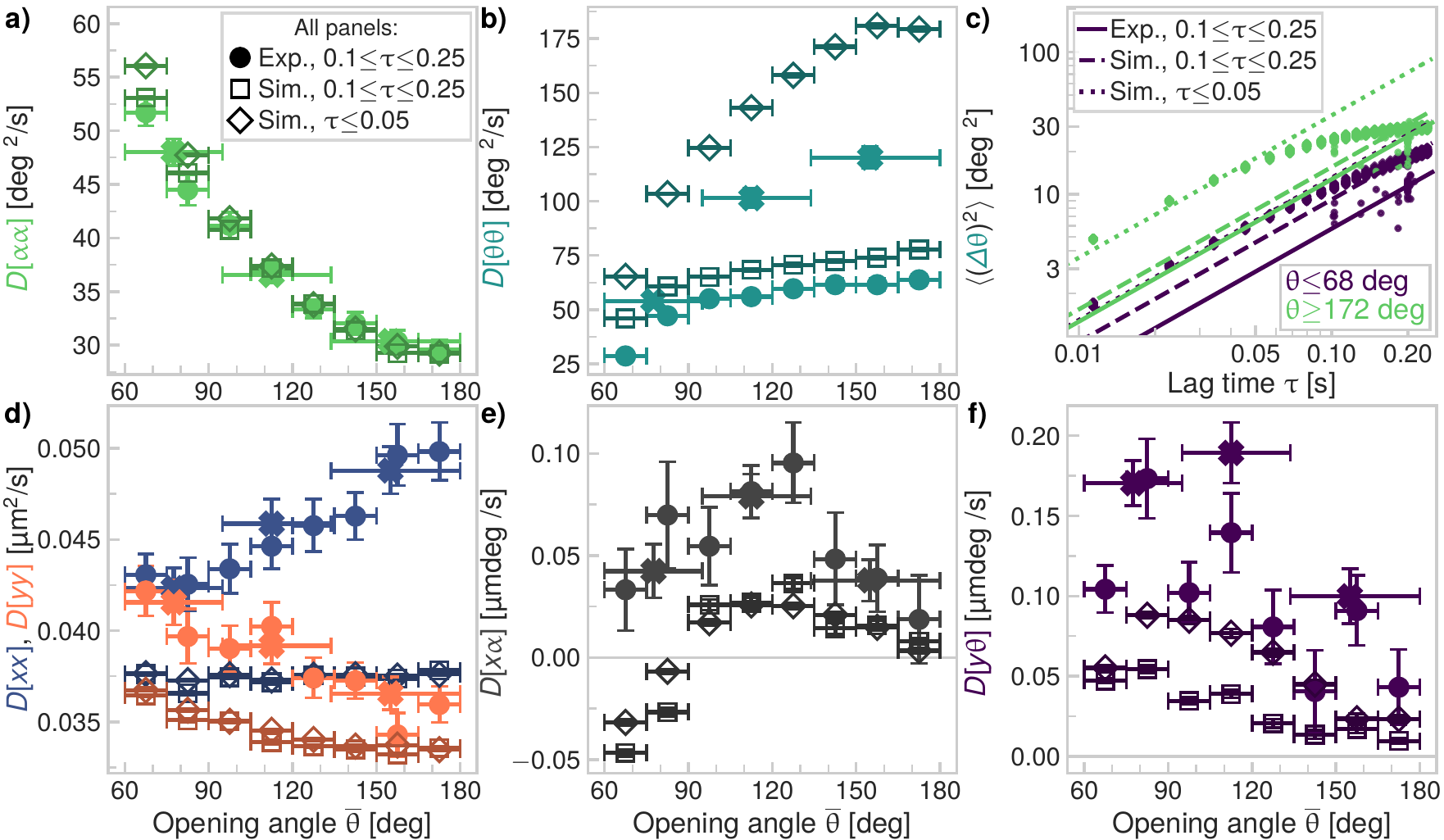}
    \caption[Trimer: effect of varying the bin size / lag time]{\textbf{Trimer: effect of varying the bin size / lag time} Comparison between $\bullet$
        experimental data, \includegraphics{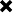} experimental data with a larger bin width (both $0.1 \leq \tau \leq \SI{0.25}{\second}$) and simulated
data: $\diamond$ $\tau \leq \SI{0.05}{\second}$, $\Box$ $0.1 \leq \tau \leq
\SI{0.25}{\second}$. \textbf{a)} The rotational diffusivities are largely
unaffected by the different choices for lagtimes (except for small opening
angles), the experimental data agrees with the simulated data. \textbf{b)} The
flexibility is highly sensitive for the choice of lagtimes. \textbf{c)}
Mean-squared angular displacement of the opening angle reveals caging effects
at longer lagtimes, which are more pronounced for higher flexibilities, an
effect inherent to the analysis method. \textbf{d)} The translational
diffusivities are less sensitive to the choice of lagtimes. The experimental
translational diffusivities are larger than the simulated ones. \textbf{e)}
Translation-rotation coupling terms. \textbf{f)} Translation-flexibility
coupling terms, including the Brownian quasiscallop mode
$D[y\theta]$~\cite{Verweij2020}, which is sensitive to the choice of
lagtimes. \label{fig:trimer_lagtimes}}
\end{figure*}

The simulations allow us to probe the diffusivity at arbitrarily high frame
rates and thus arbitrarily short lag times $\tau$, which is the time delay
between the pairs of frames considered in the calculation of the mean squared
displacements. There is a marked effect of lag time on the flexibility, as
shown in \autoref{fig:trimer_lagtimes}b. For the simulated data, we show the
results for a lag time of \SI{0.05}{\second} (diamonds) and 0.1 to
\SI{0.25}{\second} (squares, same lag time as experimental data). The lag times
of the experimental data range from 0.1 to \SI{0.25}{\second}, as set by the
frame rate of the camera. The simulated data with longer lag times are close to
the experimental data.  However, when we analyze the simulated data using a
shorter lag time, we find a large increase in the flexibility. This difference
can easily be explained by considering the mean squared angular displacement of
the opening angle in \autoref{fig:trimer_lagtimes}c.  Especially for the larger
opening angles, we see that the mean squared displacements show a plateau at
longer lag times, leading to a smaller apparent flexibility when the data is
fitted using a linear model. The effect of lag time is also present in the
Brownian quasiscallop mode in \autoref{fig:trimer_lagtimes}e.  For terms not
directly related to flexibility, such as translational diffusivity in
\autoref{fig:trimer_lagtimes}d, rotational diffusivity in
\autoref{fig:trimer_lagtimes}a and translation-rotation coupling in
\autoref{fig:trimer_lagtimes}e, we see there is no appreciable effect of
different lag times.

This plateauing for flexibility-related diffusion terms is caused by the
calculation method of the shape-dependent diffusivity. That is, we consider
only those pairs of frames where the shape of the particle stays within the
limits of the particular opening angle bin of the first frame. Therefore, if
the flexibility is high, a large percentage of frames will exceed the initial
bin and these will not be considered in the analysis. The frames where the bin
is not exceeded, as a result, are those in which the flexibility is lower,
which leads to the apparent decrease in flexibility at longer lag times. 

To solve this, larger bin widths can be used at the expense of a lower
resolution in opening angle. We tested this in \autoref{fig:trimer_lagtimes}b
and found that indeed, the values for the flexibility were higher, while the
other diffusion tensor elements were not affected (see
\autoref{fig:trimer_lagtimes}, crosses).  In fact, by using a larger bin width,
we measure the ``true'' short-time flexibility: the ratio between the
experimental flexibility for the smaller bins (circles in
\autoref{fig:trimer_lagtimes}b) and the simulated flexibility for the
experimental lag times (squares in \autoref{fig:trimer_lagtimes}b) is equal to
\num{0.78\pm0.07}. By using the larger bins, the ratio between the experimental
flexibility and the simulated flexibility at short lag times (diamonds in
\autoref{fig:trimer_lagtimes}b) is also equal to \num{0.77\pm0.07}. 

In conclusion, the diffusivity of the c.d. is very close to that relative to
the c.m. Interestingly, the interplay of lag time and bin width has a critical
effect on the measured short-time diffusion tensor elements related to shape
changes and should be carefully considered in the analysis of experimental
data.

\section{Shape-averaged coupling terms as function of chain length}

\begin{figure}
    \centering
    \includegraphics{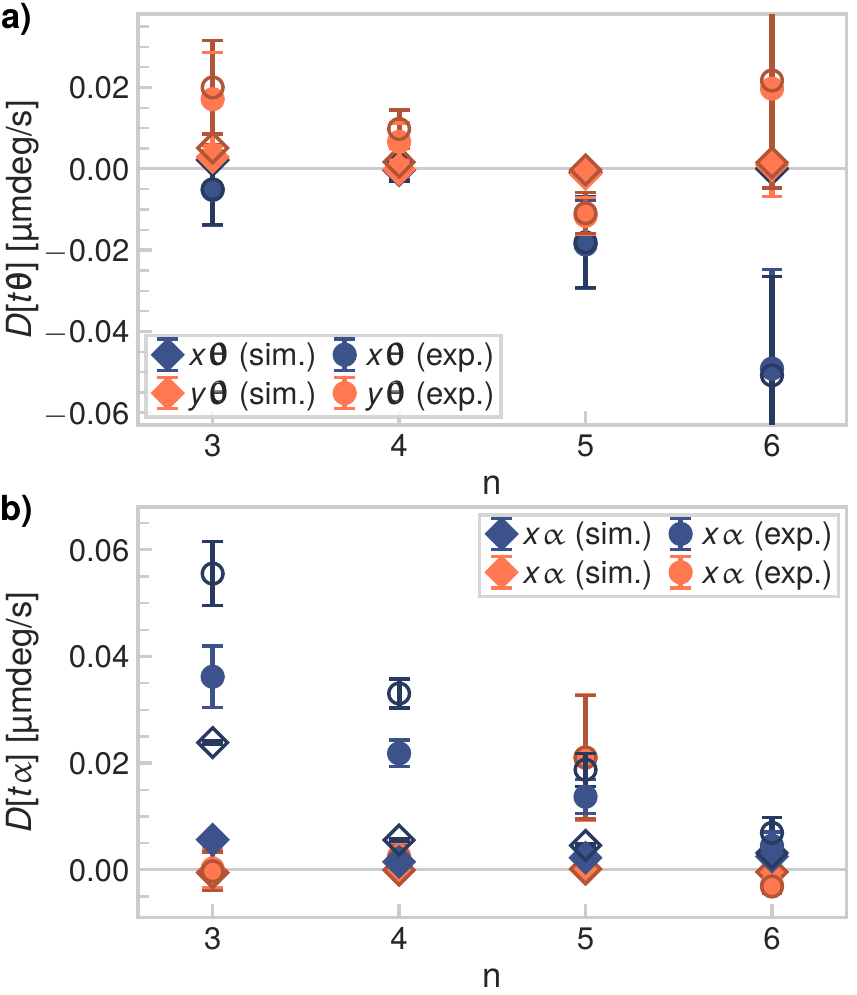}
    \caption[Shape-averaged coupling terms as function of chain length]{\textbf{Shape-averaged coupling terms as function of chain length.} In all panels, filled symbols are with respect to the c.d. and unfilled symbols are with respect to the c.m. \textbf{a)} The shape-averaged translation-flexibility coupling modes in the $y$-direction are positive for a trimer (Brownian quasiscallop mode) but average out for larger chain lengths. \textbf{b)} The shape-averaged translation-rotation coupling mode in the $x$-direction is positive for trimers and decreases as function of chain length. \label{fig:shape_avg_coupling}}
\end{figure}

We have calculated the shape-averaged translation-flexibility and
translational-rotational coupling terms with respect to both the c.m. and the
c.d., as shown in \autoref{fig:shape_avg_coupling}. The shape-averaged
translation-flexibility coupling modes in the $y$-direction are positive for a
trimer, which corresponds to the Brownian quasiscallop mode~\cite{Verweij2020}
of the shape-dependent diffusion tensor, as shown in
\autoref{fig:shape_avg_coupling}a. For longer chain lengths, this coupling term
is averaged out, most likely because of the negative correlation we have found
between the flexibility of the two opening angles of the tetramer chain, which
indicates that such coupling terms may be present for longer chain lengths as
well. Because the coordinate system is not centered on the trimer segment,
overall shape changes are taken into account. We find that overall, there is no
average coupling between translational diffusivity in the $y$-direction and the
overall flexibility $D[\bm{\theta\theta}]$.  

On the contrary, in \autoref{fig:shape_avg_coupling}b, we see that there is a
positive coupling between rotational diffusivity and translational diffusivity
in the $x$-direction, which decreases as function of chain length. This is the
same coupling we have found for the shape-dependent diffusion tensor of both
trimers~\cite{Verweij2020} and tetramers.

\bibliography{references.bib}

\end{document}